\documentclass[%
 reprint,
notitlepage,
superscriptaddress,
nofootinbib,
nobibnotes,
 amsmath,amssymb,
 aps,
floatfix,
]{revtex4-2}

\usepackage{graphicx}% Include figure files
\usepackage{dcolumn}% Align table columns on decimal point
\usepackage{bm}% bold math
\usepackage{amsmath,amssymb} 
\usepackage{amsfonts}
\usepackage{subfigure}
\usepackage{graphicx}
\usepackage{ulem}
\usepackage{color}
\usepackage{xcolor}
\usepackage{float}
\usepackage{makecell}
\usepackage{newtxtext,newtxmath}
\usepackage{mathrsfs}
\usepackage{gensymb}
\usepackage{multirow}
\usepackage{booktabs}
\usepackage{threeparttable}
\usepackage{ulem}
\usepackage{rotating}
\usepackage{graphicx}
\usepackage{tabularx}
\usepackage{aas_macros}
\usepackage{verbatim}
\usepackage{url }

\usepackage{hyperref}% add hypertext capabilities
\usepackage[mathlines]{lineno}% Enable numbering of text and display math
\begin{document}

\preprint{APS/123-QED}

\title{Accurate Kappa Reconstruction Algorithm for masked shear catalog (AKRA)}
% The last word “Algorithm” was deleted

% Force line breaks with \\

\author{Yuan Shi}
\email{yshi@sjtu.edu.cn}
\affiliation{Department of Astronomy, School of Physics and Astronomy, Shanghai Jiao Tong University, Shanghai, 200240, China}
\affiliation{Key Laboratory for Particle Astrophysics and Cosmology (MOE) / Shanghai Key Laboratory for Particle Physics and Cosmology, China}

\author{Pengjie Zhang}
\email{zhangpj@sjtu.edu.cn}
\affiliation{Department of Astronomy, School of Physics and Astronomy, Shanghai Jiao Tong University, Shanghai, 200240, China}
\affiliation{Division of Astronomy and Astrophysics, Tsung-Dao Lee Institute, Shanghai Jiao Tong University, Shanghai, 200240, China}
\affiliation{Key Laboratory for Particle Astrophysics and Cosmology (MOE) / Shanghai Key Laboratory for Particle Physics and Cosmology, China}

\author{Zeyang Sun}
\affiliation{Department of Astronomy, School of Physics and Astronomy, Shanghai Jiao Tong University, Shanghai, 200240, China}
\affiliation{Key Laboratory for Particle Astrophysics and Cosmology (MOE) / Shanghai Key Laboratory for Particle Physics and Cosmology, China}

\author{Yihe Wang}
\affiliation{Department of Astronomy, School of Physics and Astronomy, Shanghai Jiao Tong University, Shanghai, 200240, China}
\affiliation{Key Laboratory for Particle Astrophysics and Cosmology (MOE) / Shanghai Key Laboratory for Particle Physics and Cosmology, China}

\date{\today}% It is always \today, today,

\begin{abstract}
Weak gravitational lensing is an invaluable tool for understanding fundamental cosmological physics. An unresolved issue in weak lensing cosmology is to accurately reconstruct the lensing convergence $\kappa$ maps from discrete shear catalog with survey masks, which the seminal Kaiser-Squire (KS) method is not designed to address. 
    We present the Accurate Kappa Reconstruction Algorithm for masked shear catalog (AKRA) to address the issue of mask. AKRA is built upon the prior-free maximum likelihood mapmaking method (or the unbiased minimum variance linear estimator). It is mathematically robust in dealing with mask, numerically stable to implement, and practically  effective in improving the reconstruction accuracy.  Using simulated maps with mask fractions ranging from 10\% to 50\% and various mask shapes, we demonstrate that AKRA outperforms KS at both the map level and summary statistics such as the auto power spectrum $C_\kappa$ of the reconstructed map, its cross-correlation coefficient $r_\ell$ with the true $\kappa $ map, the  scatter plot and the localization measure. Unlike the Wiener filter method, it adopts no priors on the signal power spectrum, and  therefore avoids the Wiener filter related biases at both the map level and cross-correlation statistics. If we only use the reconstructed map in the unmasked regions, the reconstructed $C_\kappa$ is accurate to $1\%$ or better and $1-r_\ell \lesssim 1\%$, even for extreme cases of mask fraction and shape. As the first step, the current version of AKRA only addresses the mask issue and  therefore ignores complexities such as curved sky and inhomogeneous shape measurement noise. AKRA is capable of dealing with these issues straightfowrardly, and will  be addressed in the next version. 

% \keywords{Gravitational lensing: weak — Methods: linear mapping — Large-scale structure of Universe —
% Cosmology: observations}

% \begin{description}
% Secondary publications and information retrieval purposes.
% You may use the \texttt{description} environment to structure your abstract;
% use the optional argument of the \verb+\item+ command to give the category of each item. 
% \end{description}
\end{abstract}

\maketitle

\section{Introduction} \label{sec:intro}

Weak gravitational lensing is a powerful probe of fundamental cosmological physics such as the nature of dark matter, dark energy and gravity, and astrophysics such as the halo abundance/profile and galaxy-halo connection, to further refine our understanding of cosmology (see Ref.~\onlinecite{Bartelmann2001,Refregier2003,Munshi2008,Kilbinger2015} for reviews).  Ongoing stage III surveys, such as the Dark Energy Survey (DES \cite{Abbott2016}), the Hyper Suprime-Cam survey (HSC \cite{Aihara2018}), and the Kilo-Degree survey (KiDS \cite{Kuijken2015}), have achieved a high signal-to-noise ratio (S/N) of $\gtrsim  30$ in cosmic shear measurement. Notably, upcoming stage IV surveys, such as the China Space Station Telescope (CSST \cite{Gong2019,Yao2023CSST}), Euclid \cite{Laureijs2011},  Rubin Observatory (previously referred to as the Legacy Survey of Space and Time (LSST \cite{LSST2009})), and the Wide-Field Infrared Survey Telescope (WFIRST) (also known as Roman Space Telescope \cite{Spergel2015}) aim to achieve unprecedented precision in cosmic shear. These surveys are expected to increase the S/N of cosmic shear measurement by more than an order of magnitude, reaching S/N over 400 \citep{Yao2023CSST}.

% ,Zhang2007,Mandelbaum2018,Jullo2019,Blake2020}
%Cosmological analysis based on weak  %lensing has mainly relied on two-point %statistics of the shear field %\citep{Schneider2002}, which capture %the entire information content of a %Gaussian random field. 
Cosmic shear cosmological analysis so far relies heavily on the two-point correlation function \citep{Martin2013, Troxel2018, Hikage2019, Hamana2020, Asgari2021, Secco2022, LiXiangchong2023} or the power spectrum \citep{Hikage2011, Heymans2012, Lin2012, Kohlinger2017, Hikage2019, Nicola2021, Camacho2021}. 
%These measures provide information %about the statistical properties of %the shear field, which can be compared %with theoretical predictions and thus %infer useful cosmological constraints. 
Along with the increasing S/N, new statistics and new applications of cosmic shear are actively investigated.
The non-Gaussian nature of the lensing fields implies that there is valuable information presented in higher-order statistics or non-Gaussian statistics, such as peaks \citep{ Hilbert2012, Shan2014, Shan2018, Lin2015, Martinet2018, Peel2018, Ajani2020, Martinet2021B, Martinet2021A, Emma2022, Liu2023}, higher-order moments \citep{VanWaerbeke2013, Petri2015, Vicinanza2016, Vicinanza2018, Chang2018, Peel2018}, and Minkowski functionals \citep{Kratochvil2012, Petri2015, Vicinanza2019, Parroni2020, Zurcher2021}. 
In addition to employing these statistics, there has been growing interest in applying convolutional neural networks (CNN) to analyze the simulated weak gravitational lensing maps \citep{Gupta2018,Ribli2019a,Ribli2019b,Fluri2019,Fluri2022,Matilla2020,LuTianhuan2023}.  A significant amount of additional information is expected to extract from these statistics  \citep{Liu2015, Liu2015B, Martinet2018, Harnois2021,2023arXiv230112890E}.

%  Ongoing stage III weak lensing surveys (DES, HSC and KiDS) have achieved $S/N\ga 30$ in cosmic shear auto-correlation measurements.  Upcoming stage IV surveys (CSST, Euclid, Rubing/LSST and Roman/WFIRST) will significantly improve the S/N of cosmic shear measurement to $\ga 400$ {\color{red} ZPJ: cite some references including Yao Ji's recent paper}. 

Many of these statistics such as peak/void counts and Minkowski functionals require cosmic convergence maps, instead of cosmic shear catalogs. Convergence maps also make the measurement of higher-order correlation functions more straightforward.  So an important task is the lensing convergence map-making with discrete shear catalog. This is  complicated by issues such as the survey mask including the survey boundary, inhomogeneous measurement noise, curved sky and computational issues associated with matrix operation of large size. Among these complexities, the survey mask is a major  issue to be resolved. We propose to  resolve the mask issue by our Accurate Kappa Reconstruction Algorithm for masked shear catalog (AKRA). 
 
The seminal Kaiser Squire (KS) method \citep{Kaiser1993}, further improved by \citet{Bartelmann1995,Kaiser1995,Schneider1995,Squires1996}, performs reconstruction of the convergence field from the shear field by a linear inversion between the two fields in the ideal case. However, two primary drawbacks of the commonly used Kaiser-Squires algorithm are improper accounting for noise and survey mask effects on the shear fields and a non-local Kaiser-Squires transformation which requires knowledge of the entire sky's shear field. To overcome these limitations, many of these approaches include the forward modeling of the shear field from the convergence field via the KS transformation, and the introduction of a Bayesian prior over the convergence field \citep{Alsing2016,Alsing2017,Porqueres2022}. Proposed techniques include Wiener filtering \citep{Jeffrey2018}, sparsity priors \citep{Leonard2014,Price2019,Jeffrey2018}, log-normal priors \citep{Fiedorowicz2022b,Fiedorowicz2022a}, wavelet-based methods \citep{Starck2006,Starck2021} and others  \citep[e.g.][]{Kansal2023}. 
Various machine learning-based approaches are now being utilized to reconstruct the convergence field, e.g. Generative Neural Network (GNN) \citep{Shirasaki2021} and U-Net \citep{Jeffrey2020a}.

% \cite{Pires2020} developed the improved Kaiser Squire (KS+) method and 
% \cite{Kansal2023} also extended KS+ to the sphere (SKS+).

In this work, we investigate how to accurately recover the convergence map from observable shear maps without losing information in the presence of masks. The relationship between shear and convergence can be viewed as a linear inversion problem in Fourier space, where the survey mask can be represented as an element production that is comparable to Fourier space convolution. A plausible solution to this challenge is a widely-used deconvolution algorithm in cosmic microwave background observations \citep{Tegmark1997} and wide-field interferometric imaging. For synthesized interferometric mapmaking, images obtained by mapping visibilities and convolving them with the array synthesized beam are commonly called dirty images.
% The CLEAN deconvolution algorithm is commonly used to deconvolve these dirty images. However, in 21 cm precision cosmology (see Ref.~\onlinecite{Liu2020} for reviews), the main focus is on the faint diffuse emission, which is not effectively corrected using the CLEAN method \citep{Hogbom1974,Clark1980,Cornwell2008,Rau2011}. 
Linear deconvolution \citep{Zheng.etal2016,Dillon2020,Xu2022,Shi.etal2022a}, which is performed through matrix operations, can be done theoretically, without distinguishing between point sources and extended emission. 
The action the interferometer measurement can be described by a relationship between the discretized sky $\mathbf{x}$, and the measured visibility $\mathbf{y}$, expressed as: 
\begin{equation}
    \bf{y} = \mathbf{A}\bf{x} + \bf{n}.
\end{equation}
Here $\bf{A}$ represents the interferometric measurement operator, and $\bf{n}$ is the thermal noise.
Benefits of this approach include a data product in the image domain that potentially covers the full celestial sphere, full knowledge of the point-spread function in all directions, and thorough knowledge of the covariance matrix relating map pixels. 

To focus on the mask issue, we apply AKRA to the generation of weak lensing convergence maps in the scenario of a flat sky. In principle, this method can be applied to any survey geometry, including the entire sky case. Furthermore, we ignore shear measurement noise.  These neglected complexities can be taken into account by our method straightforwardly and will be addressed in future work.
The paper is structured as follows.  
In section \ref{sec:method}, we will present a short review of KS method (Sec. \ref{subsec:ks}), then have a detailed derivation of AKRA algorithm (Sec. \ref{subsec:lm}).  Sec. \ref{sec:simul} will cover the testing of this algorithm on simulated shear maps, as well as comparisons to the KS algorithm. Finally, Section \ref{sec:result} provides a summary of the results and discusses future directions including extending  to curved sky with inhomogeneous shape measurement noise. 

\section{Method} \label{sec:method}
In this article, we will discuss how to reconstruct the convergence from the shear map, using the Fourier space formalism.  In this section, we will first introduce the KS algorithm \citep{Kaiser1993}, which is the most widely used method to reconstruct the convergence from the shear map. Then we will introduce our AKRA algorithm, which can deal with the masked pixels and boundary effects in the shear map.

In order to provide better clarity regarding the physical quantities and related symbols involved in the reconstruction process, a detailed table has been created and is presented in Table \ref{tab:table1}. In addition to the physical quantities, the table \ref{tab:table1} also includes the mask function and the convolution kernel from the mask that is utilized in the reconstruction process. Boldface letters have been utilized to denote data vectors and matrices in the table.
The table also lists the relevant sections in which each quantity is firstly discussed.

\begin{table*}[htbp]
\begin{ruledtabular}
\centering
\caption{The physical quantities and related symbols involved in reconstruction process.}
\label{tab:table1}
\begin{tabular}{lll}

\textbf{Quantity} & \textbf{Symbol}  &\textbf{Description}  \\
\hline
\hline
Shear & $\gamma_{i}(\theta)$ (Sec.~\ref{subsec:ks}) & Two components $\gamma_1$ and $\gamma_2$ of the shear on the flat sky. \\

Shear in Fourier space & $\tilde{\gamma}_{i}(\vec{\ell})$ (Sec.~\ref{subsec:ks}) & The Fourier transform of the shear. \\

Convergence	 & $\kappa (\theta)$ (Sec.~\ref{subsec:ks})	& The convergence on the flat sky. \\

% Convergence in Fourier space & $\kappa (\vec{\ell})$ (Sec.~\ref{subsec:ks}) & The Fourier transform of the convergence. \\

% Lensing potential & $\Phi$ (Sec.~\ref{subsec:ks})& The lensing potential. \\

% Lensing potential in Fourier space & $\Phi(\vec{\ell})$ (Sec.~\ref{subsec:ks})& The Fourier transform of the lensing potential. \\

Polar angle & $\phi_{\ell}$ (Sec.~\ref{subsec:ks})& The polar angle of the Fourier mode $\vec{\ell}$. \\

Fourier mode & $\vec{\ell}$ (Sec.~\ref{subsec:ks})& The Fourier mode. \\

E-mode & $E(\vec{\ell})$ (Sec.~\ref{subsec:ks})& The E-modes components. \\

B-mode & $B(\vec{\ell})$ (Sec.~\ref{subsec:ks})& The B-modes components. \\

Unmasked convergence in Fourer space & $\tilde{\kappa}(\vec{\ell})$ (Sec.~\ref{subsec:lm}) & The Fourier transform of the unmasked convergence. \\

Mask function & $m(\theta)$ (Sec.~\ref{subsec:lm}) & The mask on the flat sky. \\

Masked shear & $\gamma_{i}^{m}(\theta)$ (Sec.~\ref{subsec:lm}) & The masked shear on the flat sky. \\

Masked shear in Fourier space & $\tilde{\gamma}_{i}^{m}(\vec{L})$ (Sec.~\ref{subsec:lm}) & The Fourier transform of the masked shear. \\

\hline
Convolution kernel from mask & $\bf{M}$ (Sec.~\ref{subsec:lm}) & The convolution kernel from mask with shape ($N_{\ell}^2, N_{\ell}^2$). \\

Observed shear vector & $\boldsymbol{\gamma}$ (Sec.~\ref{subsec:lm}) & The observed shear vector, with length $2N_{\ell}^2$. \\

Real convergence vector & $\boldsymbol{\kappa}$ (Sec.~\ref{subsec:lm}) & The real convergence vector, with length $N_{\ell}^2$. \\

Noise vector & $\boldsymbol{n}$ (Sec.~\ref{subsec:lm}) & The noise vector, with length $2N_{\ell}^2$. \\

Convolution kernel matrix & $\bf{A}$ (Sec.~\ref{subsec:lm}) & \makecell[l]{The kernel matrix transforms the convolution operation into a matrix multiplica-\\tion operation with shape ($2N_{\ell}^2, N_{\ell}^2$). It includes information from the 
mask as \\well as $\cos(2\phi_{\ell})$ and $\sin(2\phi_{\ell})$ terms.} \\

Hermitian conjugate of a matrix & $\bf{A}^{\rm{T}}$ (Sec.~\ref{subsec:lm}) & Denotes the Hermitian conjugate of matrix A.\\

Inverse of a matrix & $\bf{A}^{-1}$ (Sec.~\ref{subsec:lm}) & Denotes the inverse of matrix A.\\

Pseudo-inverse of a matrix & $\bf{A}^{+}$ (Sec.~\ref{subsec:lm}) & Denotes the Penrose-Moore pseudo-invers of matrix A.\\

Regularization matrix & $\bf{R}$ (Sec.~\ref{subsec:lm}) & The regularization matrix, which is proportional to the identity matrix.\\

Reconstructed convergence & $\boldsymbol{\kappa}^{\rm{\bf R}}$ (Sec.~\ref{subsec:lm}) & Denotes the reconstructed convergence in Fourier space.\\

Reconstructed convergence & $\kappa^{\mathrm{rec}}$ (Sec.~\ref{sec:method})& Denotes the reconstructed convergence map in real space.\\

\hline

Slope & $s$ (Sec.~\ref{sec:simul}) & \makecell[l]{The slope of a linear regression model, which is used to estimate the distribution\\ of true ($\kappa^{\rm{true}}$) and reconstructed ($\kappa^{\rm{rec}}$) convergence map in kernel density estima-\\te plot.}\\

Pearson correlation coefficient (PCC) & $\rho$ (Sec.~\ref{sec:simul}) & The PCC between $\kappa^{\rm{true}}$ and $\kappa^{\rm{rec}}$.\\

\multirow{2}*{The localization measure} & \multirow{2}*{$L$ (Sec.~\ref{sec:simul})} & A quantitative assessment of the degree to which the residuals are concentrated \\
~ & ~ & within the masked regions.\\ 

Cross correlation coefficient & $r({\ell})$ (Sec.~\ref{sec:simul}) & The cross correlation coefficient between $\kappa^{\rm{true}}$ and $\kappa^{\rm{rec}}$.\\
% \hline
\end{tabular}
\end{ruledtabular}
\end{table*}

\subsection{The Kaiser and Squires (KS) Algorithm} \label{subsec:ks}
Both the convergence and the two components of shear, $\gamma_{1}(\theta)$ and $\gamma_{2}(\theta)$, are linear combinations of the second derivatives of the lensing potential, $\Phi$. Thus, the relations between these quantities in Fourier space are linear. 

In Fourier space, they can be expressed as:
\begin{equation}
    \tilde{\gamma}_{1}(\vec{\ell})=\tilde{\kappa}(\vec{\ell}) \cos \left(2 \phi_{\ell}\right),
    \label{eq:gamma1_2}
\end{equation}

\begin{equation}
    \tilde{\gamma}_{2}(\vec{\ell})=\tilde{\kappa}(\vec{\ell}) \sin\left(2 \phi_{\ell}\right).
    \label{eq:gamma2_2}
\end{equation}
where $\sim$ denotes the Fourier transform, $\vec{\ell}$ is the wave vector in Fourier space, and $\phi_{\ell}$ is the polar angle. The convergence, $\tilde{\kappa}(\vec{\ell})$ is defined as the weighted projected matter density contrast, and can be expressed in terms of the Fourier transform of the gravitational potential.

% \begin{equation}
%     \left[\begin{array}{l}\gamma_{1}(\vec{\ell}) \\ \gamma_{2}(\vec{\ell})\end{array}\right]=\left[\begin{array}{c}\cos \left(2 \phi_{\ell_{1}}\right) \\ \sin \left(2 \phi_{\ell_{1}}\right)\end{array}\right] \cdot \tilde{\kappa}\left(\vec{\ell}\right).
%     \label{eq:gamma12_1}
% \end{equation}

% Using:
% \begin{equation}
%     \left[\begin{array}{l}\cos \left(2 \phi_{\ell_{1}}\right) \\ \sin \left(2 \phi_{\ell_{1}}\right)\end{array}\right] \cdot \left[\begin{array}{c}\cos \left(2 \phi_{\ell_{1}}\right) \\ \sin \left(2 \phi_{\ell_{1}}\right)\end{array}\right]^{T} = 1 .
% \end{equation}

% Then the equation \ref{eq:gamma12_1} can be inverted to obtain:

Using these relations in Eq.~\ref{eq:gamma1_2} and Eq.~\ref{eq:gamma2_2}, we can express the convergence as a convolution of the shear with an appropriate kernel function Kaiser and Squires \citep{Kaiser1993}, and the convergence can be expressed as:
\begin{equation}
    \tilde{\kappa}\left(\vec{\ell}\right) = \left[\begin{array}{c}\cos \left(2 \phi_{\ell_{1}}\right) \\ \sin \left(2 \phi_{\ell_{1}}\right)\end{array}\right]^{T} \cdot \left(\begin{array}{l}\tilde{\gamma}_{1}(\vec{\ell}) \\ \tilde{\gamma}_{2}(\vec{\ell})\end{array}\right).
\end{equation}
It also represents the E-mode components of the convergence map. 
 Due to the scalar nature of the lensing potential $\Phi$, weak gravitational lensing produces only E modes in the shear field. The KS inversion is an ideal estimator of the convergence map in the absence of masks or holes in the data, but it can introduce spurious features or artifacts when applied to realistic data sets with incomplete sky coverage. In this paper, we modify the above formalism and propose a new algorithm to deal with the masked shear field.

% By combining the equations for $\tilde{\gamma}_{1}(\vec{\ell})$ and $\tilde{\gamma}_{2}(\vec{\ell})$, we also get the E-modes and B-modes components of the convergence map. The E-mode can be expressed as:
% \begin{equation}
% \tilde{E}(\vec{\ell})=\tilde{\gamma}_{1}(\vec{\ell}) \cos \left(2 \phi_{1}\right)+\tilde{\gamma}_{2}(\vec{\ell}) \sin \left(2 \phi_{\ell}\right)=\tilde{\kappa}(\vec{\ell}).
% \label{eq:Emode}
% \end{equation}

% And the B-mode is :
% \begin{equation}
% \tilde{B}(\vec{\ell})=\tilde{\gamma}_{1}(\vec{\ell}) \sin \left(2 \phi_{1}\right)-\tilde{\gamma}_{2}(\vec{\ell}) \cos \left(2 \phi_{\ell}\right)=0.
% \label{eq:Bmode}
% \end{equation}

\subsection{Accurate Kappa Reconstruction Algorithm for masked shear catalog (AKRA) algorithm} \label{subsec:lm}
% The Kaiser and Squires algorithm in above section \ref{subsec:ks}
% is based on the assumption that the shear field is fully observed, then we can directly obtain the convergence field by inverting the shear field. 
% In standard cosmology, the convergence sourced by a real density field should be a pure E-mode. 

We first define a mask function $m(\vec{\theta})$ in real space, which is equal to 1 for the observed region and 0 for the masked region. Then the observed shear field can be expressed as:
\begin{equation}
    \gamma_{i}^m(\vec{\theta})=m(\vec{\theta})\gamma_{i}(\vec{\theta}),
    \label{eq:gamma_obs}
\end{equation}
where the mask function $m(\vec{\theta})$ and the shear field $\gamma_i(\vec{\theta})$ are both defined in real space, and their shapes are ($N_{\theta}, N_{\theta}$). 

Then in Fourier space, the shear field can be expressed as the convolution of the mask function and the true shear field:
\begin{equation}
    \begin{aligned} \tilde{\gamma}_{i}^m(\vec{L}) & =\int \gamma_{i}(\vec{\theta}) m(\vec{\theta}) e^{-i \vec{L} \cdot \vec{\theta}} d^{2} \theta \\ & =\int \frac{d^{2} \ell_{1}}{(2 \pi)^{2}} \int d^{2} \ell_{2} \tilde{\gamma}_{i}\left(\vec{\ell}_{1}\right) \tilde{m}\left(\vec{\ell}_{2}\right) \delta^{D}\left(\vec{\ell}_{1}+\vec{\ell}_{2}-\vec{L}\right) \\ & \propto \int d^{2} \vec{\ell}_{1} \tilde{\gamma}_{i}\left(\vec{\ell}_{1}\right) \tilde{m}\left(\vec{L}-\vec{\ell}_{1}\right).
    \end{aligned}
    \label{eq:gamma_convolution}   
\end{equation}
Here, $\tilde{\gamma}_i\left(\vec{\ell}_1\right)$ and $\tilde{m}\left(\vec{\ell}_2\right)$ are the Fourier transforms of the true shear and mask functions respectively, and its shape is ($N_{\ell}, N_{\ell}$). 
This equation reveals that the observed shear field is a convolution of the true shear field and mask function. The Fourier space convolution operation is a crucial step in reconstructing the convergence field from the shear map.
Then the integral can be discretized as a sum over Fourier modes,
\begin{equation}
    \begin{aligned} \tilde{\gamma}_{i}^m(\vec{L}) &= \sum_{\vec{\ell}_{1}=1}^{N_{\ell}^2} \tilde{\gamma}_{i}\left(\vec{\ell}_{1}\right) \tilde{m}\left(\vec{L}-\vec{\ell}_{1}\right)\Delta \Omega  \\ &=\sum_{\vec{\ell}_{1}=1}^{N_{\ell}^2} \tilde{\gamma}_{i}\left(\vec{\ell}_{1}\right) M\left(\vec{L}, \vec{\ell}_{1}\right) \Delta\Omega.
    % \\
    % &= \mathbf{M} \cdot \tilde{\gamma}_{i}\left(\vec{\ell}_{1}\right), 
    \end{aligned} 
    \label{eq:gamma_convolution2}
\end{equation}
Here $\Delta \Omega$ is the pixel area in Fourier space, $M\left(\vec{L}, \vec{\ell}_{1}\right) = \tilde{m}\left(\vec{L}-\vec{\ell}_{1}\right)$ is the convolution kernel fuction from the mask, and its shape is ($N_{\ell}^2, N_{\ell}^2$).

Eq.~\ref{eq:gamma_convolution2} can be expressed reasonably well in the form of matrix multiplication:
\begin{equation}
    \tilde{\gamma}_{1}^m(\vec{L}) = \mathbf{M} \cdot \tilde{\gamma}_{1}(\vec{\ell}_{1}), \quad
    \tilde{\gamma}_{2}^m(\vec{L}) = \mathbf{M} \cdot \tilde{\gamma}_{2}(\vec{\ell}_{1}). 
    \label{eq:gamma_obs2}
\end{equation}
In Eq.~\ref{eq:gamma_obs2}, the true shear field $\gamma_{i}(\vec{\ell_{1}})$ can be reshaped as a vector with length $N_{\ell}^2$, whereas the matrix $\mathbf{M}$ is a convolution kernel with a shape of ($N_{\ell}^2, N_{\ell}^2$). Using this equation, the convolution operation between shear and mask in Fourier space can be simplified to a single matrix multiplication operation.

Substitute Eq.~\ref{eq:gamma1_2} and Eq.~\ref{eq:gamma2_2} into the above Eq.~\ref{eq:gamma_obs2}, we can get the masked shear field in Fourier space:
\begin{equation}
    \begin{array}{l}
        \tilde{\gamma}_{1}^m(\vec{L}) = \mathbf{M} \cdot \tilde{\kappa}\left(\vec{\ell}_{1}\right) \cos \left(2 \phi_{\ell_{1}}\right),  \\ 
        \tilde{\gamma}_{2}^m(\vec{L})=\mathbf{M} \cdot \tilde{\kappa}\left(\vec{\ell}_{1}\right) \sin \left(2 \phi_{\ell_{1}}\right).
    \end{array}
\end{equation}
Here the $\tilde{\kappa}\left(\vec{\ell}_{1}\right)$ is a vector with length $N_{\ell}^2$, and $\tilde{\kappa}\left(\vec{\ell}_{1}\right) \cos \left(2 \phi_{\ell_{1}}\right)$ means multiplying each element of the vector $\tilde{\kappa}\left(\vec{\ell}_{1}\right)$ by $\cos \left(2 \phi_{\ell_{1}}\right)$. Then we can rewrite the above equation as:
\begin{equation}
    \left[\begin{array}{l}\tilde{\gamma}_{1}^m(\vec{L})\\ \tilde{\gamma}_{2}^m(\vec{L})\end{array}\right]=
    \left[\begin{array}{c}\cos \left(2 \phi_{\ell_{1}}\right) \mathbf{M} \\ \sin \left(2 \phi_{\ell_{1}}\right) \mathbf{M} \end{array}\right] \cdot \tilde{\kappa}\left(\vec{\ell}_{1}\right),
    \label{eq:gamma_obs3}
\end{equation}
where $\cos \left(2 \phi_{\ell_{1}}\right)$ and $\sin \left(2 \phi_{\ell_{1}}\right)$ are both vectors with length $N_{\ell}^2$, so we can multiply each element in $\cos \left(2 \phi_{\ell_{1}}\right)$ (or $\sin \left(2 \phi_{\ell_{1}}\right))$ with each row of the matrix $\mathbf{M}$, and then stack the results together to form a new matrix $\bf{A}$ with a shape of $2N_{\ell}^2 \times N_{\ell}^2$. 

In Eq.~\ref{eq:gamma_obs3}, the data from shear maps are linearly related to the convergence field. Grouping the two components of the shear field together, we can get a new vector $\tilde{\gamma}^m(\vec{L})$ with length $2N_{\ell}^2$, and then we can rewrite Eq.~\ref{eq:gamma_obs3} as:
\begin{equation}
    \boldsymbol{\gamma}=\mathbf{A} \boldsymbol{\kappa}+\boldsymbol{n}.
    \label{eq:gamma_linear}
\end{equation}

With a total of $2N_{\ell}^2$ observation points from $\boldsymbol{\gamma}$, $\bf{A}$ is a $2N_{\ell}^2  \times N_{\ell}^2$ matrix,
$\boldsymbol{\kappa}$ is a $N_{\ell}^2$-element vector with each element corresponding to each Fourier mode of the real convergence map. Meanwhile, the noise vector, $\boldsymbol{n}$, has dimensions of $2N_{\ell}^2$. For a Gaussian random noise model, with $\langle\boldsymbol{n}\rangle=0$ and variance of $\sigma_{\boldsymbol{n}}^2$, the covariance matrix for the noise is given by $ \bf{N} \equiv\left\langle\boldsymbol{n} \boldsymbol{n}^{\rm{T}}\right\rangle$. Here we set $\mathbf{N}^{-1} = \mathbf{I}$ for noise-free case.

Our analysis, which focuses on the weak lensing shear field, has primarily addressed the mask effect but has not extensively considered the errors associated with real measurement. Several factors, including uncertainties in the point spread function and intrinsic alignment (IA), contribute to shear measurement errors. These errors lead to correlated noise, rendering the noise covariance matrix $\mathbf{N}$ non-diagonal. This matrix can be incorporated within the AKRA framework, albeit in a modified $\mathbf{N}$ matrix form. 
To mitigate the effects of the point spread function and IA, the strategies proposed by \citet{Giblin2021, Gatti2021} and \citet{Yao2023A&A} also present effective alternatives.
Shape noise persists as the predominant source of contamination in mass mapping, leading to inhomogeneous noise. By applying a smoothing kernel to an adequate number of galaxies, as indicated by \citet{Liu2015jul}, we anticipate that the residual shape noise, after smoothing, will approximate a Gaussian distribution, pursuant to the central limit theorem.
Further complexities arise from sub-pixel errors due to the discrepancy between the sky's infinite resolution and the fixed resolution of our maps, a challenge also prevalent in Cosmic Microwave Background (CMB) mapmaking \cite{Dunner2013, Qu2024ApJ, Naess2023}. Given these considerations, the covariance matrix $ \bf{N}$ of the noise is non-identity in real observations. Future studies should investigate these effects to improve mass mapping analysis accuracy and reliability.

    % \begin{aligned}
    %     \hat{\boldsymbol{\kappa}}&=\left(\mathbf{A}^{\rm{T}} \mathbf{N}^{-1} \mathbf{A}\right)^{-1} \mathbf{A}^{\rm{T}} \mathbf{N}^{-1} \boldsymbol{\gamma}\\
    %     &=\mathbf{D}\mathbf{A}^{\rm{T}} \mathbf{N}^{-1} \boldsymbol{\gamma}. \\
    % \end{aligned}

To optimally estimate $\boldsymbol{\kappa}$, we use the minimum variance estimator \citep{Tegmark1997}:
\begin{equation}
    \hat{\boldsymbol{\kappa}} = \mathbf{D}\mathbf{A}^{\rm{T}} \mathbf{N}^{-1} \boldsymbol{\gamma}. 
    \label{eq:kappa_estimator}
\end{equation}
where $\mathbf{D}$ is some invertible normalization matrix.
To analyze the statistics of this estimator, the mean and covariance can be calculated. Since the thermal noise has $\langle\boldsymbol{n}\rangle=0$,
we can write the ensemble average of the estimator as:
\begin{equation}
    \begin{aligned}\langle\hat{\boldsymbol{\kappa}}\rangle & =\left\langle\mathbf{D} \mathbf{A}^{\mathrm{T}} \mathbf{N}^{-1} \boldsymbol{\gamma}\right\rangle \\ & =\mathbf{D}\left(\mathbf{A}^{\mathrm{T}} \mathbf{N}^{-1} \mathbf{A}\right) \boldsymbol{\kappa} \\ & \equiv \mathbf{P}\boldsymbol{\kappa},\\\end{aligned}
\end{equation}
where $\mathbf{P}=\mathbf{D}\left(\mathbf{A}^{\mathrm{T}} \mathbf{N}^{-\mathbf{1}} \mathbf{A}\right)$ is the matrix-valued PSF.

% The quality of the reconstructed map is a vital consideration in the reconstruction process. The point spread function (PSF) and the estimator's covariance can be used to quantify the quality. 
% The expectation of the estimator can be expressed as: 
The covariance of the estimator is
\begin{equation}
    \mathbf{C} \equiv\left\langle(\hat{\boldsymbol{\kappa}}-\boldsymbol{\kappa})(\hat{\boldsymbol{\kappa}}-\boldsymbol{\kappa})^{\mathrm{T}}\right\rangle=\mathbf{P}\mathbf{D}^{\mathrm{T}}.
    \label{eq:covariance}
\end{equation}

More generally the estimators of $\boldsymbol{\kappa}$ can be formed with different choices of $\bf{D}$. If we wanted an unbiased estimator of the sky, we should choose $\mathbf{D} = \left(\mathbf{A}^{\rm{T}} \mathbf{N}^{-1} \mathbf{A}\right)^{-1} $ to obtain idealized PSF $\bf{P} = \bf{I}$. Then Eq. \ref{eq:covariance} becomes
\begin{equation}
    \mathbf{C}=\mathbf{P}\mathbf{D}^{\mathrm{T}} = \left(\mathbf{A}^{\rm{T}} \mathbf{N}^{-1} \mathbf{A}\right)^{-1}.
    \label{eq:covariance2}
\end{equation}
Then the inverse covariance matrix is $\mathbf{C}^{-1} = \mathbf{A}^{\rm{T}} \mathbf{N}^{-1} \mathbf{A}$. This implies that $\mathbf{A}^\dagger\mathbf{N}^{-1}\mathbf{A}$ measures the information content in our maps. In Sec. \ref{subsec:random_mask}, we will explore the structure of this matrix for different cases of mask.

In practice, inverse problems related with imaging are often ill-posed. For example, in the case of interferometric data reconstruction, the matrix $\bf{A}^{\rm{T}} \bf{N}^{-1}\bf{A}$ is often numerically non-invertible due to the instrument's insensitivity to certain linear combinations of the sky. \citet{Shi.etal2022a} proposed to use a pseudo-inverse matrix (the Moore-Penrose pseudo-inverse):
\begin{equation}
    \mathbf{D} \sim\left(\mathbf{A}^{\mathrm{T}} \mathbf{N}^{-\mathbf{1}} \mathbf{A}\right)^{+}.
\end{equation}

\citet{Zheng.etal2016} also demonstrated that a regularization matrix $\mathbf{R}$ could be added in 
Eq. \ref{eq:kappa_estimator} as follows:
\begin{equation}
    \hat{\boldsymbol{\kappa}}^{\bf{R}}=\left(\mathbf{A}^{\rm{T}} \mathbf{N}^{-1} \mathbf{A} +\bf{R}\right)^{-1} \mathbf{A}^{\rm{T}} \mathbf{N}^{-1} \boldsymbol{\gamma}.
    \label{eq:kappa_estimator2}
\end{equation}
We set the deconvolution matrix $\mathbf{D} = \left(\mathbf{A}^{\rm{T}} \mathbf{N}^{-1} \mathbf{A} +\bf{R}\right)^{-1}$, where $\bf{R}$ is a diagonal regularization matrix with the same size as $\bf{A}^{\rm{T}} \bf{N}^{-1}\bf{A}$. We choose $\bf{R} = \varepsilon \mathbf{I}$, with $\varepsilon$ being a small number that depends on $\mathbf{A}^{\rm{T}} \mathbf{N}^{-1} \mathbf{A}$ and ensures numerical stability. Since the maximum eigenvalue of $\mathbf{A}^{\rm{T}} \mathbf{N}^{-1} \mathbf{A}$ is 1, we set $\varepsilon = 10^{-4}$ to minimize the numerical error. The PSF matrix is
\begin{equation}
    \mathbf{P}=\left(\mathbf{A}^{\rm{T}} \mathbf{N}^{-1} \mathbf{A} +\bf{R}\right)^{-1}\left(\mathbf{A}^{\mathrm{T}} \mathbf{N}^{-1} \mathbf{A}\right).
    \label{eq:psf}
\end{equation}

In this paper, we utilize the method of regularization, as presented in Eq.~{\ref{eq:kappa_estimator2}}, to calculate the estimator for $\boldsymbol{\kappa}$. 
Eq.~\ref{eq:kappa_estimator2} can also be extended to include a prior model with known uncertainties \citep{Zheng.etal2016}. However, we do not assume any prior model for the convergence map in this paper to obtain an unbiased estimator. 

To evaluate our estimator in Eq. \ref{eq:kappa_estimator2}, computational efficiency warrants careful consideration, especially given the large data volumes anticipated in forthcoming surveys. The primary computational expense within our algorithm is the construction and multiplication of matrices,  which have a computational complexity of $\mathcal{O}(N_{\rm{pix}}^2)$.
Fortunately, the process of matrix multiplication is well-suited for parallel computation. As a complementary strategy, we advocate for the segmentation of observational data into manageable flat sky patches. Additionally, employing iterative methods for linear systems, such as the conjugate gradient method \citep{Oh1999}, offers an effective alternative for handling large and sparse systems.

\begin{figure}[htbp]
    \centering
    \includegraphics[width=0.5\textwidth]{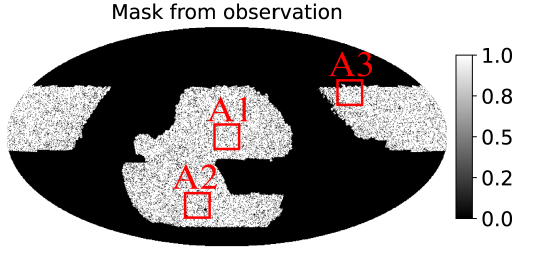}
    \caption{The mask is generated from real observations. The region in red box represents three specific masked region shown in Fig.~\ref{fig:mask_A1-3}.}
    \label{fig:mask_obs}
\end{figure}

\begin{figure*}[htbp]
    \centering
    \includegraphics[width=0.8\textwidth]{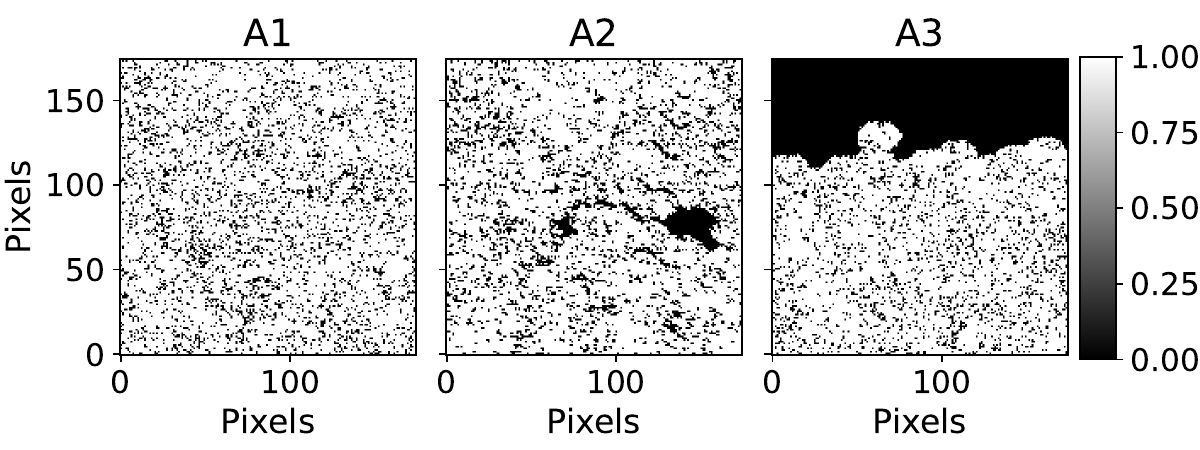}
    \caption{Three patches of sky with angular resolution of 6.7 arcmin (HEALPix Nside = 512). Left panel: position 1 (A1) at RA [$-10^{\circ}$,$10^{\circ}$], Dec. [$-10^{\circ}$,$10^{\circ}$]. Middle panel: position 2 (A2) at RA [$20^{\circ}$,$40^{\circ}$], Dec. [$-60^{\circ}$,$-40^{\circ}$]. Right panel: position 3 (A3) at RA [$-120^{\circ}$,$100^{\circ}$], Dec. [$20^{\circ}$,$40^{\circ}$]. The masked pixels are denoted in black and the unmasked pixels in white.}
    \label{fig:mask_A1-3}
\end{figure*}

\begin{figure*}[htbp]
    \centering
    \includegraphics[width=0.8\textwidth]{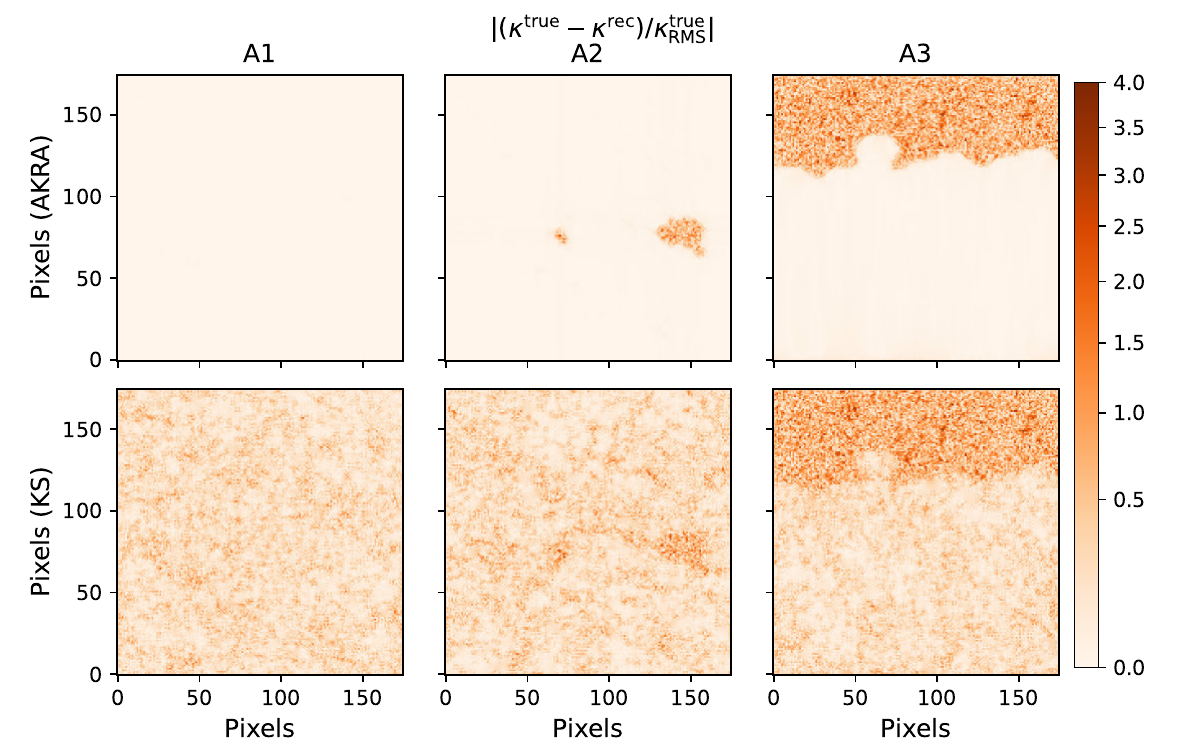}
    \caption{The residual maps normalized by r.m.s. of the true signal from AKRA method (top row) and KS method (bottom row).
    From left to right, the three columns correspond to the three patches of sky in Fig.~\ref{fig:mask_A1-3}. The KS method has large residuals due to masked pixels in all A1-3 cases. However, The AKRA method remains mostly robust, except for some regions in A2 and A3 with clustered masked pixels.}
    \label{fig:rec_obs}
\end{figure*}

\begin{figure}[htbp]
    \centering
    \includegraphics[width=0.49\textwidth]{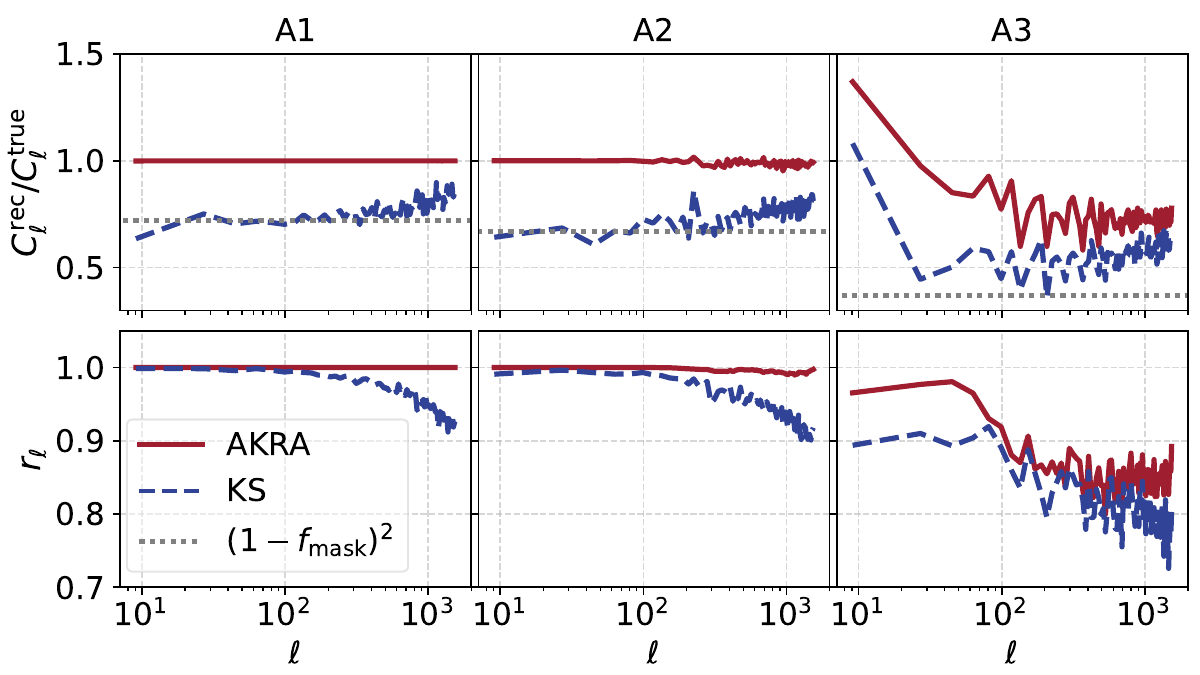}
    \caption{Results for all pixels. Top row: The power spectrum ratio between the reconstructed map and the true $\kappa$ map. Bottom row: the cross-correlation coefficient between the reconstructed and the true convergence map. From left to right, the three columns correspond to the three patches of sky in Fig.~\ref{fig:mask_obs}. AKRA can recover the power spectrum more accurately than KS method in all cases.
    AKRA also has a higher cross-correlation coefficient than KS method, indicating a better agreement between the reconstructed and the true maps.} 
    \label{fig:rec_obs_pk_all}
\end{figure}

\begin{figure}[htbp]
    \centering
    \includegraphics[width=0.49\textwidth]{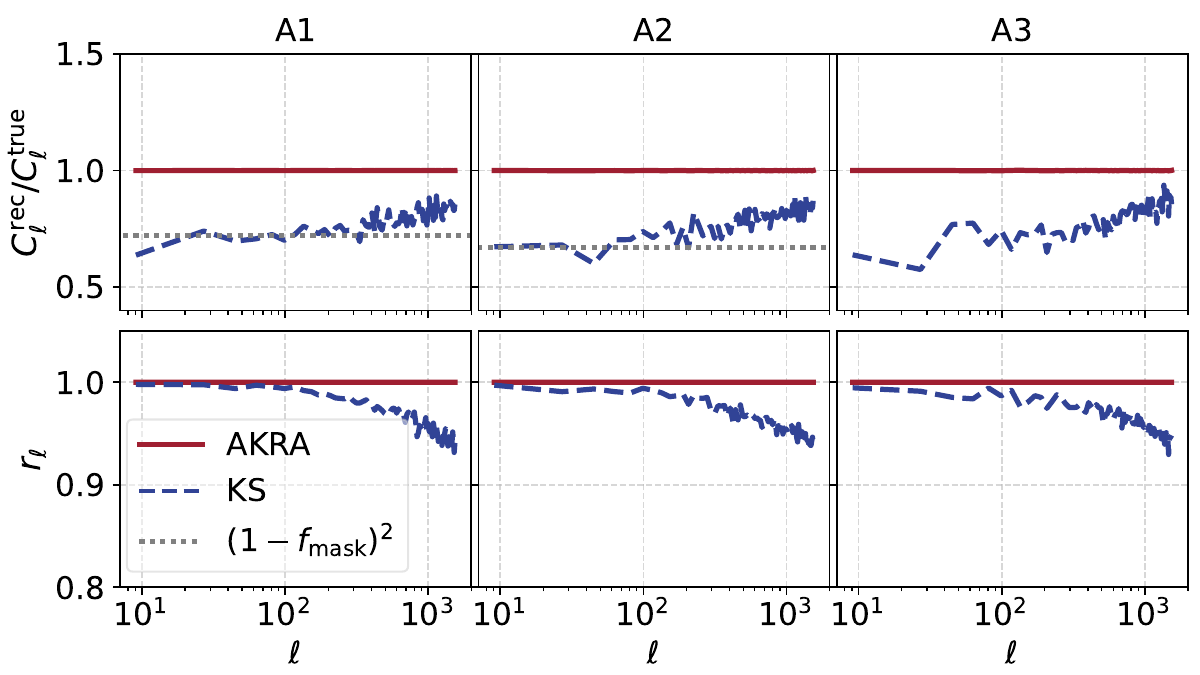}
    \caption{Same as Fig.~\ref{fig:rec_obs_pk_all}, but only for the unmasked pixels. AKRA has a significantly higher level of accuracy, with an accuracy of $1\%$ or better, and a cross-correlation coefficient of $1-r_\ell \lesssim 1\%$.} 
    \label{fig:rec_obs_pk_good}
\end{figure}

\begin{figure}[htbp]
    \centering
    \includegraphics[width=0.49\textwidth]{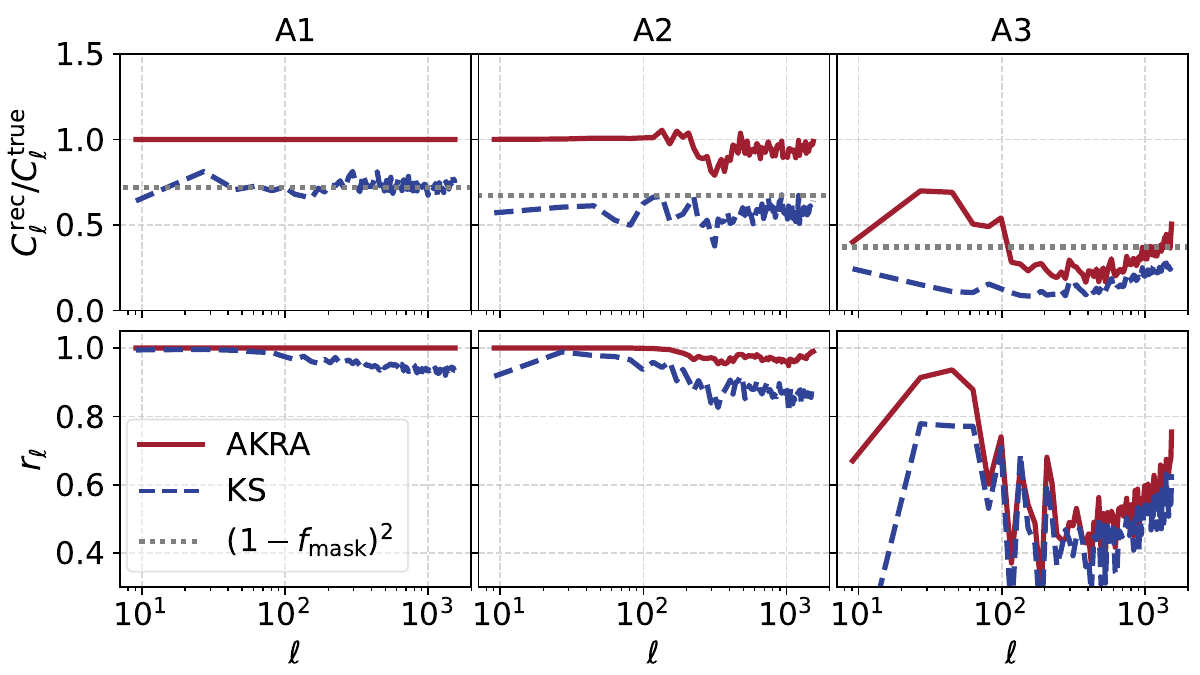}
    \caption{Same as Fig.~\ref{fig:rec_obs_pk_all}, but only for the masked pixels. For the A1 case, AKRA performs the power spectrum ratio and cross-correlation coefficient very well, with both values close to 1. For the A2 cases, AKRA shows some deviations in estimating the power spectrum ratio and cross-correlation coefficient, which are more noticeable at small scales.
     } 
    \label{fig:rec_obs_pk_bad}
\end{figure}

\begin{figure}[htbp]
    \centering
    \includegraphics[width=0.49\textwidth]{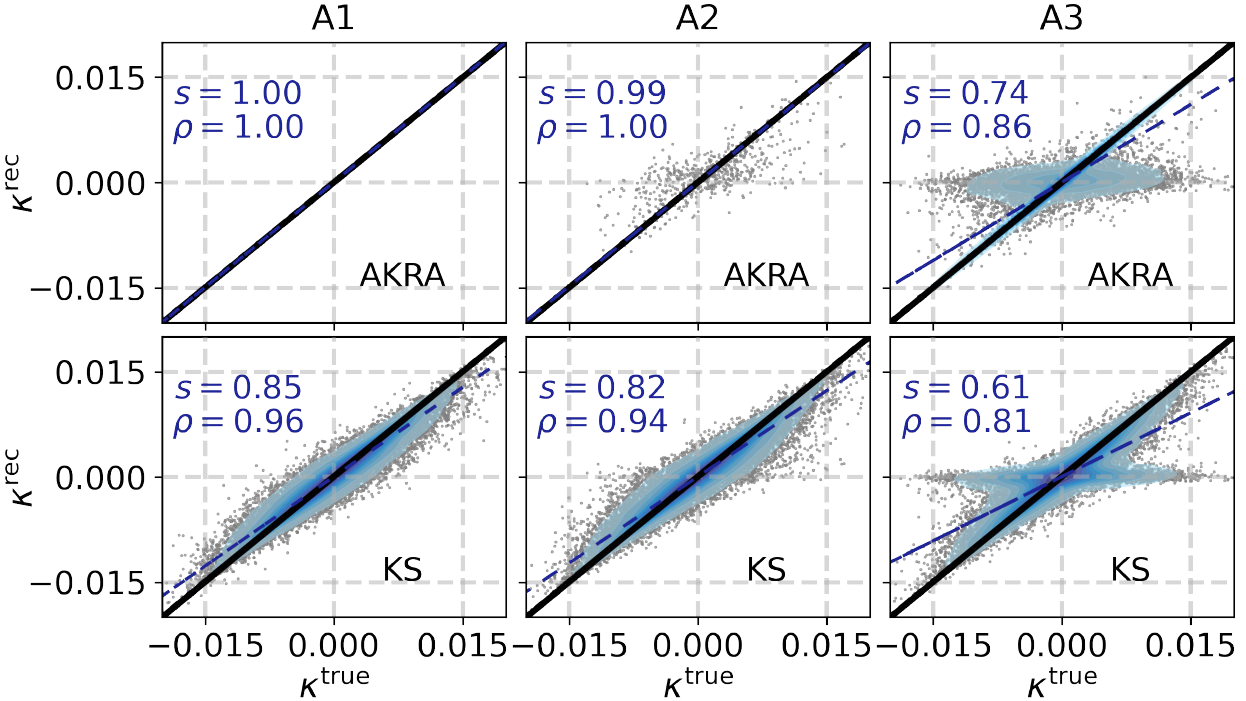}
    \caption{The $\kappa_{\rm rec}$-$\kappa_{\rm true}$ scatter plot for all pixels. From left to right, the three columns correspond to the three patches of sky in Fig.~\ref{fig:mask_obs}. The data points are displayed as gray dots, with their x-axis and y-axis denoting the pixel values for the input $\kappa$ map and the reconstructed $\kappa$ map, respectively. 
    The blue dashed line represents the result obtained from fitting a regression model of the pixels from the data points. The black solid line represents the ideal result. The slope $s$ of the blue dashed line and the PCC $\rho$ are also shown in the figure. As a result of all pixels, the recovered $\kappa$ maps are almost identical to the input $\kappa$ maps for positions A1 and A2, while the recovered $\kappa$ maps differ significantly from the input $\kappa$ maps for position A3.}     
    \label{fig:kde_obs}
\end{figure}

\begin{figure}[htbp]
    \centering
    \includegraphics[width=0.49\textwidth]{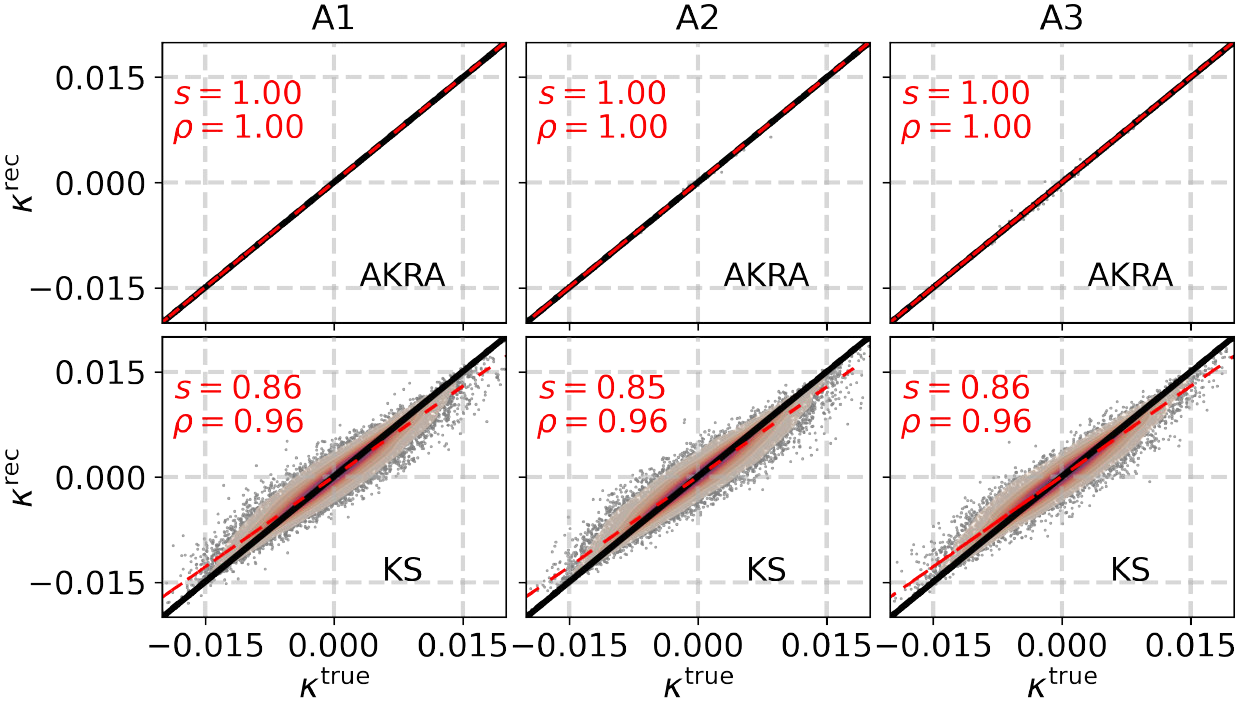}
    \caption{Same as Fig.~\ref{fig:kde_obs}, but only for the unmasked pixels. AKRA reconstruct $\kappa$ map for all cases accurately, with the slope $s$ close to 1 and the Pearson correlation coefficient $\rho$ equal to 1. In AKRA results, several points at the boundary for A3 will deviate slightly from the ideal outcome.  }
    \label{fig:kde_obs_good}
\end{figure}

\begin{figure}[htbp]
    \centering
    \includegraphics[width=0.49\textwidth]{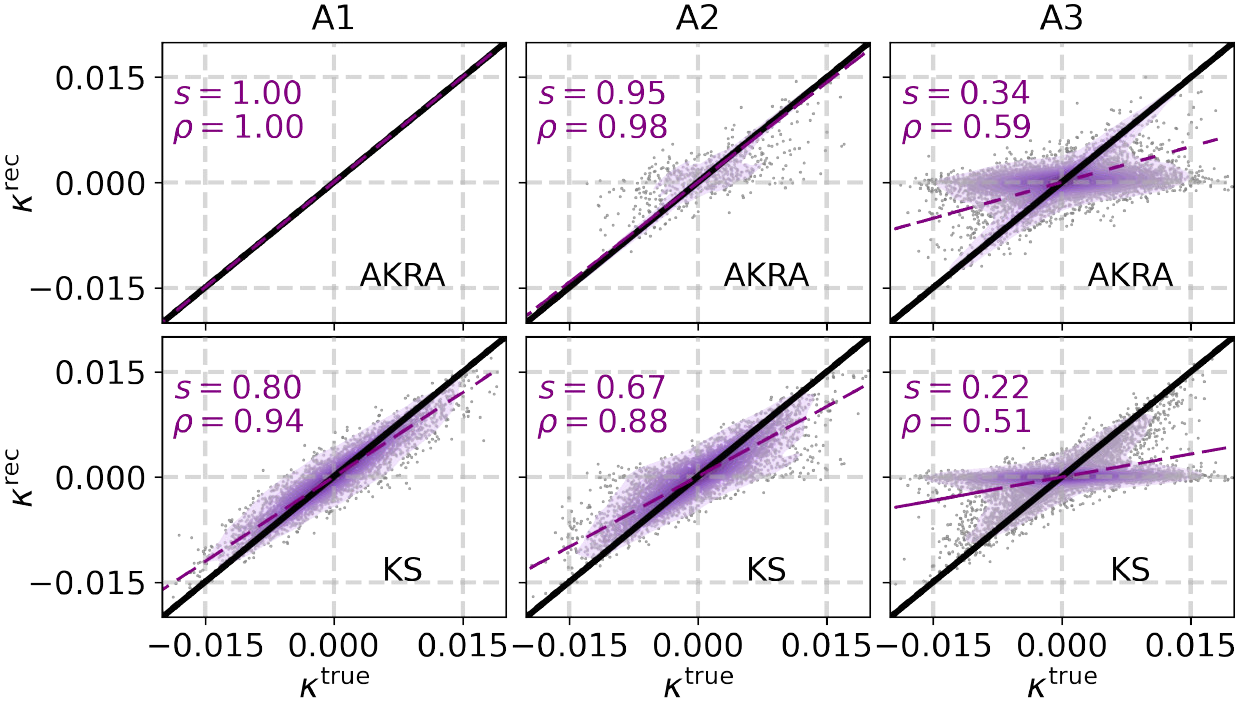}
    \caption{Same as Fig.~\ref{fig:kde_obs}, but only for the masked pixels. AKRA can only reconstruct $\kappa$ map for A1 case accurately. }
    \label{fig:kde_obs_bad}
\end{figure}

\section{Simulation} \label{sec:simul}
In this section, we present a $\boldsymbol{\kappa}$ reconstruction simulation under the assumption of a flat sky, noise-free, and periodic boundary conditions. 

The simulation process involves several stages:
\begin{enumerate}
    \item {\bf Generation of convergence field:} Use a power spectrum from cosmological parameters in Planck 2018 flat $\Lambda$CDM cosmology \citep{Planck2018} to generate a Gaussian random field. Subsequently, an inverse Fourier transform is carried out to obtain the convergence map $\kappa(\vec{\theta})$.
    \item {\bf Generation of shear fields:} From the convergence map, generate the shear fields $\gamma_{1}(\vec{\theta})$ and $\gamma_{2}(\vec{\theta})$ from the convergence map using Eq.~\ref{eq:gamma1_2} and Eq.~\ref{eq:gamma2_2} in the Fourier field, followed by an inverse Fourier transform to obtain the shear map in the real field.
    \item {\bf Adding a mask to the shear field:} The mask is added to the shear field $\gamma_{1}(\vec{\theta})$ and $\gamma_{2}(\vec{\theta})$, which results in the masked shear field $\gamma_{1}^{m}(\vec{\theta})$ and $\gamma_{2}^{m}(\vec{\theta})$. In this section, we will introduce three kinds of mask.
    \item {\bf Generation of convolution kernel matrix:} Generate the convolution kernel $\bf{M}$ matrix in Eq.~\ref{eq:gamma_convolution2} from the mask.
    \item {\bf Modification of convolution kernel matrix:} Multiply $\cos(2\phi_{\ell})$ and $\sin(2\phi_{\ell})$ terms into the matrix $\bf{M}$, and then we can get the  convolution kernel matrix $\bf{A}$ in Eq.~\ref{eq:gamma_linear}.
    \item {\bf Solving the linear equation:} Calculate the estimator $\hat{\boldsymbol{\kappa}}$ by Eq.\ref*{eq:kappa_estimator2}, and then we obtain the reconstructed convergence map in real space using an inverse Fourier transform.
\end{enumerate}

To further investigate the quality of the reconstructed maps in a more quantitative manner, we list a few commonly used statistics.

\begin{itemize}
\item The accuracy of the auto power spectrum $C^{\rm rec}_\ell$ of the reconstructed convergence $\kappa_{\rm rec}$, quantified by the ratio $C^{\rm rec}_\ell/C^{\rm true}_\ell$ as a function of multipole $\ell$. Here $C^{\rm true}_\ell$ is the auto power spectrum of the true map. This is one of the key measures of map quality for the purpose of weak lensing cosmology.
\item The cross-correlation coefficient $r_\ell$ as a function of multipole $\ell$.
\begin{equation}
    r_\ell\equiv \frac{C_{\rm rec-true}(\ell)}{\sqrt{C^{\rm rec}(\ell)C^{\rm true}(\ell)}}\ .
\end{equation}
Here $C_{\rm rec-true}(\ell)$ is the cross-power spectrum between the reconstructed and true convergence maps. $r_\ell$ quantifies the accuracy of reconstructing the phase. It is also a key measure of map quality. $r_\ell\neq 1$ will cause bias of amplitude of ${\mathcal O}(1-r_\ell)$ in cross-correlations between  $\kappa^{\rm rec}$ and other large scale structure fields. Unlike the error in the amplitude  which can be corrected at map level by scaling each $\ell$ mode with $\sqrt{C^{\rm rec}/C^{\rm true}}$ estimated from simulations,  errors in the phase  ($r_\ell\neq 1$) can not be corrected at map level. Therefore the requirement of $r_\ell=1$ is often more challenging. 
\item  The $\kappa_{\rm rec}$-$\kappa_{\rm true}$ scatter plot, also called Kernel Density Estimation (KDE). The plot is a non-parametric way to estimate the probability density function of a random variable, which is used to compare the distribution of true convergence map and reconstructed convergence map. We also define the slope $s$ as the bestfit to $\kappa_{\rm rec}$-$\kappa_{\rm true}$ by $\kappa_{\rm rec}=s\kappa_{\rm true}$. The solution for $s$ is	
    \begin{equation}
        s=\frac{\sum_i \kappa^{\rm rec}_i\kappa^{\rm true}_i}{\sum_i \kappa^{\rm true}_i\kappa^{\rm true}_i}.
    \label{eq:slope}
    \end{equation}
    By examining $s$ of the $\kappa_{\rm rec}$-$\kappa_{\rm true}$ scatter plot, we can discern the level of similarity between the two distributions. A slope of $s=1$ indicates the reconstructed convergence field exactly matches the true convergence distribution.

    \item The Pearson correlation coefficient (PCC). The PCC is a measure of the linear correlation between two convergence fields, which is defined as:
    \begin{equation}
            \rho=\frac{\left\langle\kappa^{\rm{true}} \kappa^{\rm{rec}}\right\rangle}{\sqrt{\left\langle(\kappa^{\rm{true}})^{2}\right\rangle} \sqrt{\left\langle(\kappa^{\rm{rec}})^{2}\right\rangle}}.
            \label{eq:PCC}
    \end{equation}

    The PCC is used to measure the similarity between true and reconstructed convergence map. The PCC ranges from -1 to 1. The closer the PCC is to 1, the more similar the two variables are.
    
    % \item The power spectrum and cross-correlation coefficient. The power spectrum is a measure of the distribution of the convergence field in Fourier space, which is defined as $P_{{\kappa_1} \kappa_1}=V  \kappa_1^{*}(\ell) \kappa_1(\ell)$. The cross-correlation coefficient is defined as:
    % $$
    % r({\ell})=\frac{P_{{\kappa_1} \kappa_2}}{\sqrt{P_{\kappa_1 \kappa_1} P_{\kappa_2 \kappa_2}}},
    % $$
    % where $P_{{\kappa_1} \kappa_2}$ is the cross power which is defined as:
    % $$
    % P_{{\kappa_1} \kappa_2}=V  \kappa^{*}_2(\ell) \kappa_1(\ell).
    % $$
    
    \item The localization measure, denoted as $L$, quantifies the extent to which the residuals are localized within the masked regions:
    \begin{equation}
        L\equiv \frac{\sum_i |\Delta_i|m_i}{(\sqrt{2/\pi}\sigma_{\kappa^{\rm true}})(N_{\rm pix}(1-f_{\rm mask}))}.
    \label{eq:localization}
    \end{equation}
    Here $\Delta_i\equiv \kappa^{\rm rec}_i-\kappa^{\rm true}_i$. If Gaussian $\langle |\kappa|\rangle=\sqrt{2/\pi}\sigma_\kappa$. $N_{\rm pix}$ is the total number of pixels and $f_{\rm mask}$ is the fraction of masked regions. Ideally, when $m=1$, $\Delta=0$. But since the shear-convergence relation is non-local in real space, masks in the shear catalog impact the reconstruction of convergence in the unmasked regions. Namely $L\neq 0$. The value of $L$ then serves as a measure for the localization of residuals, indicating to what extent the shear mask contaminates the convergence reconstruction in the masked regions. $L\ll 1$ is desirable. 
\end{itemize}

\begin{figure*}[htbp]
    \centering
    \includegraphics[width=0.95\textwidth]{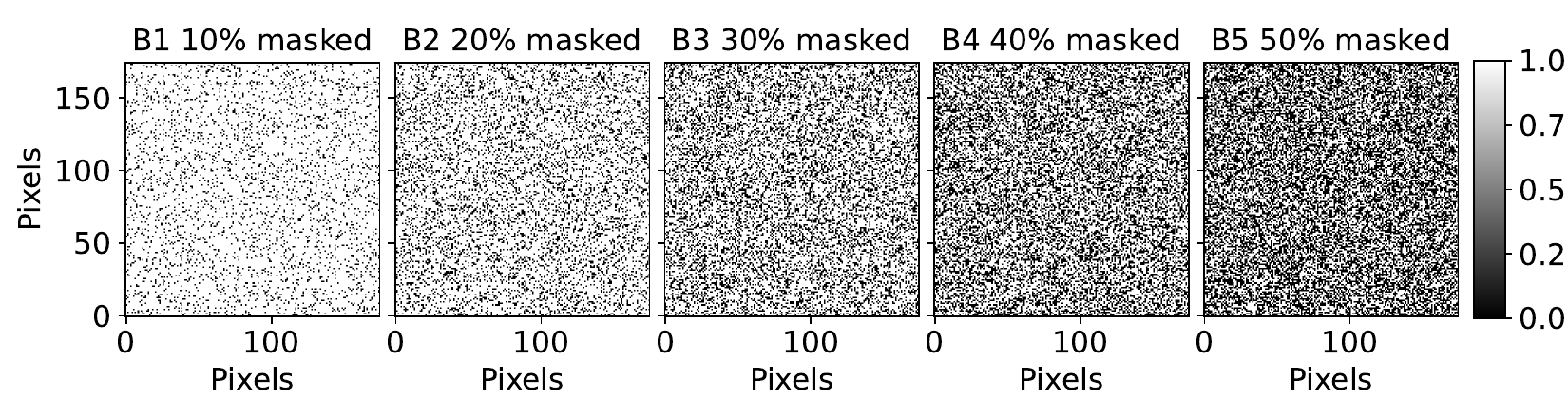}
    \caption{The mask is created by randomly selecting masked pixels from $m(\theta)$ and setting them to zero. The masks in panels B1 to B5 have masked pixel rates of 10\%, 20\%, 30\%, 40\%, and 50\%, respectively.}
    \label{fig:random_mask}
\end{figure*}

We simulate the reconstruction of convergence maps from shear maps using a flat sky, noise-free, and periodic model. In this study, we aim to discuss three types of masks that have been utilized in our simulation. The first one is obtained from real observations (see Fig. \ref{fig:mask_A1-3}), the second type of mask is randomly generated (see Fig. \ref{fig:random_mask}) with varying mask fractions, and the third one is a mask with a specific shape and fixed mask fraction (see Fig. \ref{fig:circular_mask}).

\begin{table*}[ht]
    \begin{ruledtabular}
    % \begin{threeparttable}
    \centering
    \caption{Mask used in simulation and statistics results. This table summarizes the results of three types of simulations: observation, random mask, and circular mask. The third to tenth columns show: mask fraction, PSF matrix ratio, slope $s$ of the scatter plots for $\kappa_{\rm rec}$-$\kappa_{\rm true}$, Pearson correlation coefficient $\rho$, and localization measure $L$ for both methods. Based on the results of $s$, $\rho$, and $L$, it is evident that the AKRA method outperforms the KS method in all cases.}
    \label{tab:table2}
    \begin{tabularx}{\textwidth}{cccccccccc}
    % \hline
    % \hline
    \textbf{Mask name}& \textbf{Type} \footnotemark[1] &\textbf{$f_{\text{mask}}$} \footnotemark[2] & \textbf{$\frac{\overline{\left(\rm{Off\_diag}(\bf{P})\right)^2}}{\overline{\left(\rm{Diag}(\bf{P})\right)^2}}$} \footnotemark[3]& \textbf{$s$} (KS) \footnotemark[4] & \textbf{$s$} (AKRA) & \textbf{$\rho$} (KS)  &\textbf{$\rho$} (AKRA) &\textbf{$L$} (KS)  & \textbf{$L$} (AKRA)\\
    \hline
    \hline
    A1 & Observation & 15\% & $6.69\times 10^{-13}$& 0.85 & 1.00 & 0.96 & 1.00 & 0.256 & $1.83\times 10^{-4}$ \\
    A2 & Observation & 18\% & $1.55\times 10^{-7}$ & 0.82 & 0.99 & 0.94 & 1.00 & 0.260 & $8.88 \times 10^{-4}$ \\
    A3 & Observation & 39\% & $3.51\times 10^{-6}$ & 0.61 & 0.74 & 0.81 & 0.86 & 0.258 & $4.40\times 10^{-3}$ \\
    % 0.2564835365918883 / 0.25955319076509853/ 0.2583472721497145
    %1.83e-04 %8.88e-04 %4.40e-03
    \hline
    B1 & Random & 10\% & $7.98\times 10^{-14}$ & 0.90 & 1.00 & 0.97 & 1.00 & 0.211 & $ 1.38\times 10^{-4}$ \\
    B2 & Random & 20\% & $4.77\times 10^{-13}$ & 0.80 & 1.00 & 0.94 & 1.00 & 0.333 & $ 2.17\times 10^{-4}$ \\
    B3 & Random & 30\% & $3.90\times 10^{-12}$ & 0.70 & 1.00 & 0.91 & 1.00 & 0.427 & $ 4.28\times 10^{-4}$\\
    B4 & Random & 40\% & $2.33\times 10^{-10}$ & 0.60 & 1.00 & 0.86 & 1.00 & 0.522 & $ 1.74\times 10^{-3}$ \\
    B5 & Random & 50\% & $2.51\times 10^{-7}$  & 0.50 & 0.98 & 0.81 & 0.99 & 0.611 & $ 7.73\times 10^{-2}$ \\
    \hline
    C1 & Circular & 10\% & $1.21\times 10^{-13}$ & 0.90 & 1.00 & 0.97 & 1.00 & 0.181 & $1.38\times 10^{-4}$   \\
    C2 & Circular & 10\% & $1.62\times 10^{-11}$ & 0.90 & 1.00 & 0.97 & 1.00 & 0.153 & $1.99\times 10^{-4}$   \\
    C3 & Circular & 10\% & $1.91\times 10^{-8}$  & 0.90 & 1.00 & 0.96 & 1.00 & 0.127 & $5.70 \times 10^{-4}$  \\
    C4 & Circular & 10\% & $1.54\times 10^{-7}$  & 0.90 & 0.99 & 0.96 & 0.99  & 0.112 & $8.70 \times 10^{-4}$ \\
    C5 & Circular & 10\% & $3.07\times 10^{-7}$  & 0.90 & 0.98 & 0.96 & 0.99 & 0.099 & $9.29 \times 10^{-4}$ \\
    % \hline
    \end{tabularx}
    \end{ruledtabular}

        \footnotetext[1]{The type of mask: observation (Sec.~\ref{subsec:real_mask}), random (Sec.~\ref{subsec:random_mask}) and circular (Sec. ~\ref{subsec:circular_mask}).}
        \footnotetext[2]{The percentage of masked pixels.}
        \footnotetext[3]{The ratio of the average of the squared off-diagonal elements of the PSF matrix to the average of the squared diagonal elements of the PSF matrix.}
        \footnotetext[4]{$s$ is the slope of the best-fit line to the $\kappa_{\rm rec}$-$\kappa_{\rm true}$ scatter plot, as defined by Eq.~\ref{eq:slope}. $\rho$ is the Pearson correlation coefficient between the reconstructed and true convergence maps, as given by Eq.~\ref{eq:PCC}. $L$ is the localization measure defined by Eq.~\ref{eq:localization}.
        Here the $s$ and $\rho$ values listed are calculated using all pixels (both masked and unmasked) from the convergence maps reconstructed with AKRA and KS.}
        
    % \end{threeparttable}
\end{table*}

\subsection{Mask from real observation}
\label{subsec:real_mask}
% We use Dark Energy Camera Legacy Survey (DECaLS) DR8 shear catalog to generate real observation mask. DECaLS contains galaxy images in $g, r, z$ bands in the North Galactic Cap (DECaLS-NGC) at $\delta \leq +32.375 \degree$ and the entire South Galactic Cap (DECaLS-SGC, including the data contributed by Dark Energy Survey). 
% A conventional shear calibration \citep{Heymans2012, Miller2013, Hildebrandt2017} is applied as
% \begin{equation}
%     \gamma^{\rm obs} = (1+m)\gamma^{\rm true} + c \ ,
% \end{equation}
% with a multiplicative bias $m$ and an additive bias $c$. The shear measurement and imperfect modeling of point-spread function (PSF) size result in the multiplicative bias. 
% % To calibrate our shear catalog, we cross-matched the DECaLS DR8 objects with the external shear measurements, including Canada-France-Hawaii Telescope (CFHT) Stripe 82, Dark Energy Survey  \citep[DES;][]{DES2016} and Kilo-Degree Survey \citep[KiDS;][]{Hildebrandt2017}. 
% The additive bias comes from residuals in the anisotropic PSF correction which depends on galaxy sizes, and it is subtracted from each galaxy in the catalog.
We use $5.18$ million galaxies at redshift bin $0.8<z<1.0$, which covering $\sim$ 13,000 deg$^2$. To consider the real survey geometry, we generate the binary mask from the real observation of DESI imaging surveys DR8 with nside = 1024. The mask is 1 when if are shear galaxies located at the pixel, otherwise it is 0. Then downgrade the resolution to nside = 512. The downgraded mask is 1 when pixel values $\geq 0.5$, otherwise is 0, as depicted in Fig. \ref{fig:mask_obs}.

\begin{figure}[htbp]
    \centering
    \includegraphics[width=0.45\textwidth]{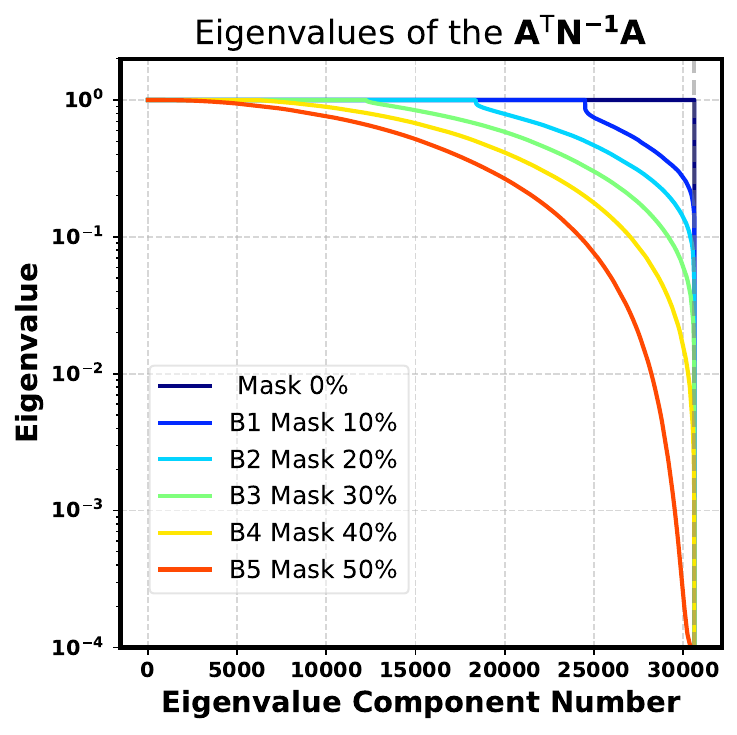}
    \caption{Eigenvalue spectrum of the matrix $\mathbf{A}^{\rm{T}} \mathbf{N}^{-1} \mathbf{A}$ for different mask maps. The magnitude of the eigenvalues of this matrix indicates the precision of the measurements for different modes in the map. It is also the matrix we would have to invert if we intend to set the PSF to the identity. The amplitude of this matrix decrease as the mask coverage increases from B1 to B5. 
    We can also measure similar independent modes with mask fractions ranging from 10\% to 50\%. This suggests that the singularity of the matrix $\mathbf{A}^{\rm{T}} \mathbf{N}^{-1} \mathbf{A}$ is not solely attributed to the mask fraction.}
    \label{fig:eigenvalue}
\end{figure}

\begin{figure}[htbp]
    \centering
    \includegraphics[width=0.45\textwidth]{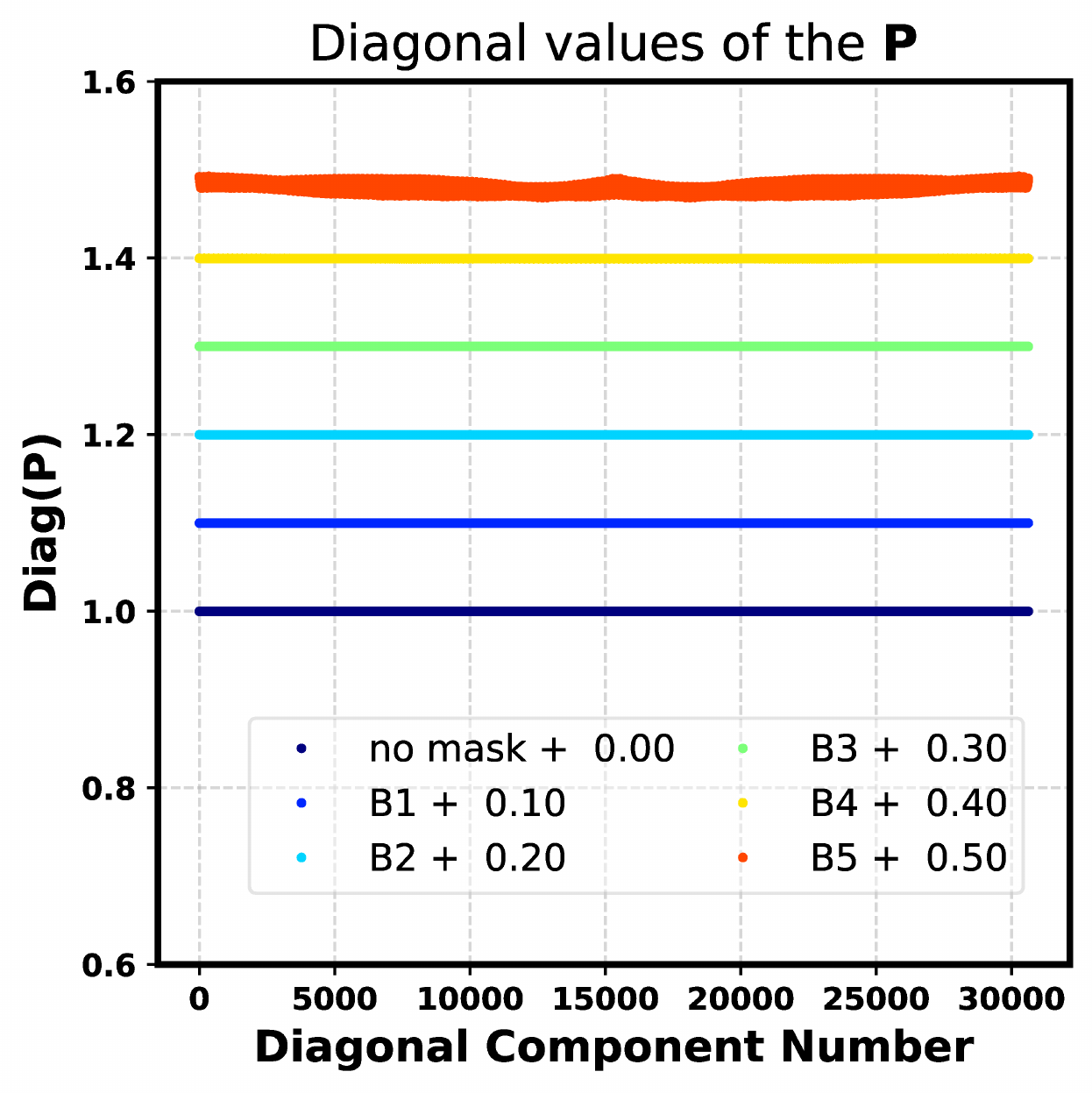}
    \caption{Diagonal elements of the PSF matrix $\mathbf{P}$ for different mask maps. A PSF matrix is optimal when it has 1 on the main diagonal and 0 elsewhere. This indicates that the pixels are well reconstructed. For cases B1 to B5, the PSF matrix is almost optimal, with diagonal values very close to 1. In case B5, the diagonal values are slightly lower than 1 and display minimal oscillation due to the presence of a large number of masked pixels.}
    \label{fig:diag}
\end{figure}

\begin{figure}
    \centering
    \includegraphics[width=0.45\textwidth]{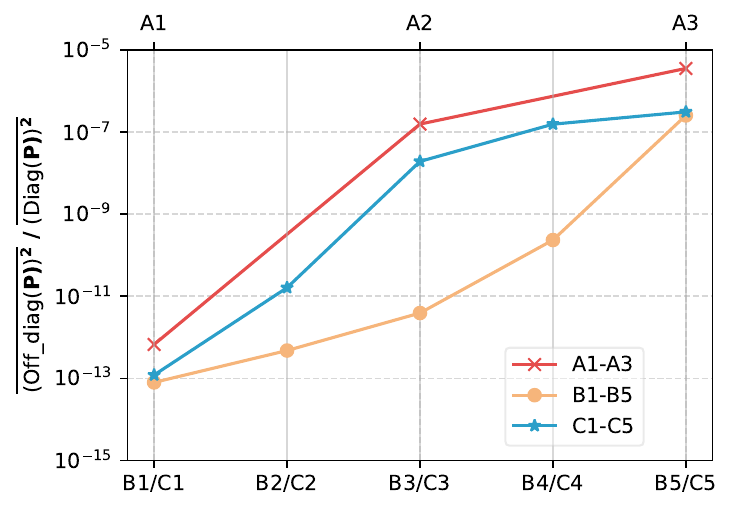}
    \caption{The ratio of squared off-diagonal elements to squared diagonal elements of the PSF matrix, corresponding to the fourth column in Table \ref{tab:table2}. In all cases, the ratio is significantly lower than 1. With increasing mask fraction or clustering, the ratio will increase.}
    \label{fig:psf_ratio}
\end{figure}

The AKRA method described in this paper will focus on a specific region of the sky.
As an example, we selected three particular patches from real observation mask in Fig.~
\ref{fig:mask_obs}, shown in Fig.~\ref{fig:mask_A1-3}. The masked pixels are denoted in black and the unmasked pixels in white. 
The masked pixels in position 1 (A1) are randomly distributed, whereas the mask in position 2 (A2) is more concentrated and results in a small area full of masked pixels. 
The mask in position 3 (A3) is a combination of the type of mask in position 1 and 2, with distinct borders. Each of the three patches of sky has $175^2$ pixels, with a side length of 20$^{\circ}$.

In this study, we followed the procedures outlined in the prior section to acquire the masked shear field and reconstruct the convergence map. Firstly, in steps 1 and 2, we generated the shear map $\gamma_{1}(\vec{\theta})$ and $\gamma_{2}(\vec{\theta})$ by converting the convergence map $\kappa(\vec{\theta})$ from the power spectrum. We then subject the shear fields $\gamma_{1}(\vec{\theta})$ and $\gamma_{2}(\vec{\theta})$ to a mask in step 3 to produce the masked shear field $\gamma_{1}^{m}(\vec{\theta})$ and $\gamma_{2}^{m}(\vec{\theta})$ for further analysis. A representation of the mask maps used in this process is visible in Fig.~\ref{fig:mask_A1-3}. Next, we reconstructed the convergence map $\kappa^{\rm{rec}}$ by applying KS method and AKRA algorithm to the masked shear field $\gamma_{1}^{m}(\vec{\theta})$ and $\gamma_{2}^{m}(\vec{\theta})$(in steps 4 to 6), respectively.

\begin{figure*}[htbp]
    \centering
    \includegraphics[width=0.8\textwidth]{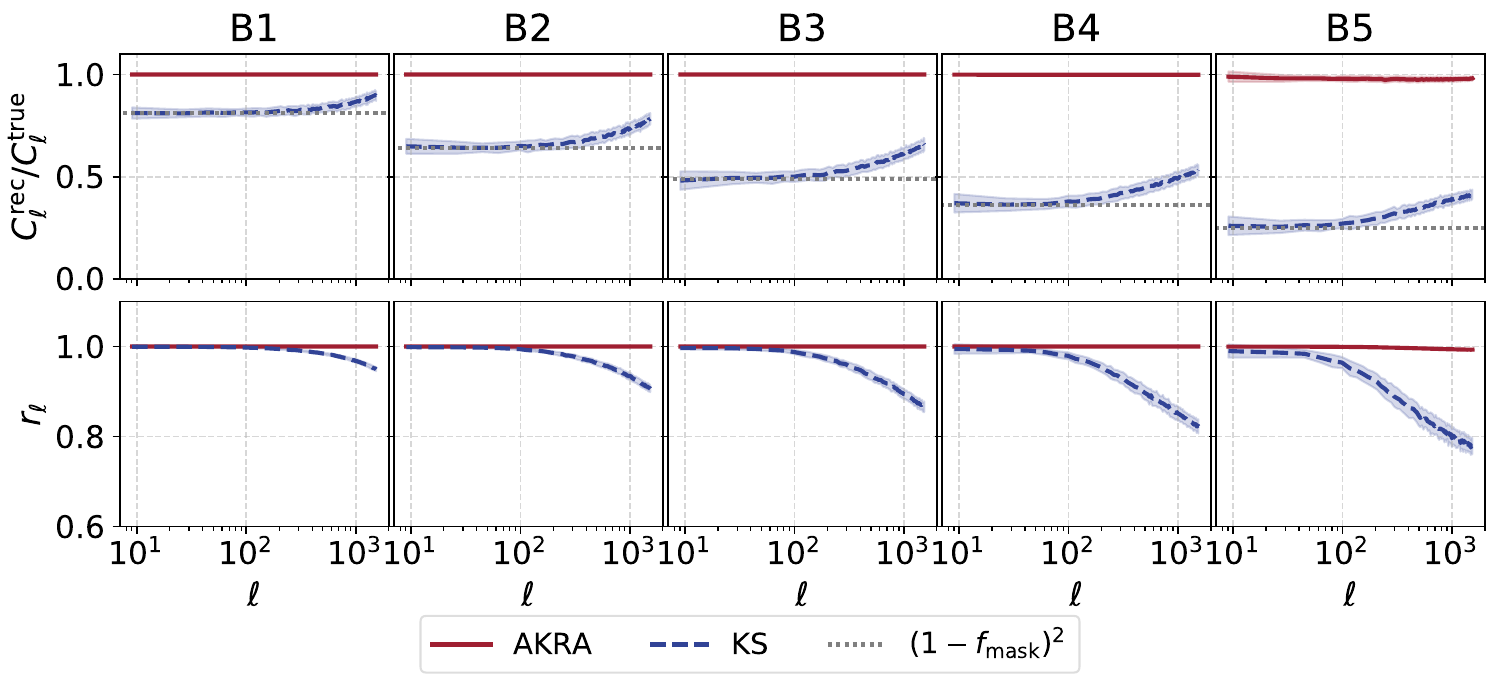}
    \caption{Results for random masks. Top: power spectrum ratio for 100 realizations (unmasked pixels). The blue and red regions represent the 1$\sigma$ confidence interval. Bottom: cross-correlation coefficient between the reconstructed and the true convergence map. Here we did not correct the bias in the power spectrum obtained from the KS method, resulting from varying mask rates, as the impact on different scales is not uniform. AKRA accurately recovers the true power spectrum in the simulation, with a power spectrum ratio and cross-correlation coefficient in all cases around $1\%$ and $1-r_\ell \lesssim 1\%$ respectively.}
    \label{fig:random_mask_ps}
\end{figure*}

Fig. \ref{fig:rec_obs} illustrates the residual maps $\kappa^{\rm{true}}-\kappa^{\rm{rec}}$ obtained from the AKRA method (top row) and the KS method (bottom row). These maps are then normalized by the root mean square (r.m.s.) of the true signal. The KS method produces significant residuals due to masked pixels in all A1-3 cases. 
The AKRA method, on the other hand, is robust to masked pixels, as evidenced by the zero residuals in the A1 case. For the A2 case, the AKRA method performs well for isolated masked pixels, but slightly deteriorates for clustered masked pixels. For the A3 case, a clear boundary is visible in the mask map, which is also reflected in the residual maps from the AKRA method. Above the boundary, where there are no observations and all pixels are masked, the normalized residual maps are mostly greater than or equal to 1, indicating that the residual is comparable to the signal.

The power spectrum ratios in Figs.~\ref{fig:rec_obs_pk_all}, \ref{fig:rec_obs_pk_good}, and \ref{fig:rec_obs_pk_bad} consistently demonstrate the superior accuracy of AKRA in recovering the power spectrum compared to the KS method. For the unmasked pixels, AKRA produces the reconstructed $C_{\ell}$ is accurate to $1\%$ or better. Surprisingly, in the case of A1, AKRA also accurately reconstructs the masked pixels, resulting in a power spectrum ratio close to 1.

The cross-correlation analysis presented in Figs.~\ref{fig:rec_obs_pk_all}, \ref{fig:rec_obs_pk_good}, and \ref{fig:rec_obs_pk_bad} further solidifies AKRA's advantages. It yields significantly higher agreement between reconstructed and true maps, demonstrating superior performance versus KS. Notably, for high-quality regions AKRA attains near-perfect coefficients of $1-r_\ell \lesssim 1\%$. Equally impressive is its ability to attain the cross-correlation coefficient of $r_{\ell}\approx1$ for masked pixels in $A1$ case, an unprecedented achievement of accuracy for masked pixels. This suggests AKRA is capable of recovering masked pixels in certain situations, owing to the non-local nature of cosmic shear observables and convergence. 

To evaluate the global quality of the maps in real space, we constructed a $\kappa_{\rm rec}$-$\kappa_{\rm true}$ scatter plot for each patch of sky, as shown in Fig.~\ref{fig:kde_obs}. The black solid line represents the ideal result (slope $s$ = 1.0). The reconstructed $\kappa$ maps generated using the AKRA method for positions A1 and A2 are almost indistinguishable from the input $\kappa$ map. However, the reconstructed $\kappa$ map for position A3 differs from the input $\kappa$ map due to the mask in this patch containing a large number of masked pixels, with a slope value of 0.74 much lower than the ideal result. Moreover, we also present the PCC $\rho$ of the pixels from both maps in Fig.~\ref{fig:kde_obs}. The coefficient $\rho$ demonstrates the same trend as the slope $s$.

Tab. \ref{tab:table2} also show the detailed results of statistics for the three patches of sky. 
The third to tenth columns show the mask fraction, the ratio of the average of the squared off-diagonal elements of the PSF matrix to the average of the squared diagonal elements of the PSF matrix, the slope $s$ of the scatter plots for $\kappa_{\rm rec}$-$\kappa_{\rm true}$, the Pearson correlation coefficient $\rho$, and the localization measure $L$ from both methods for all pixels. 
The localization measure $L$, defined in Eq. \ref{eq:localization}, quantifies the level of contamination from mask in the shear catalog on the reconstructed convergence in the unmasked regions. L of AKRA is factor of ${\mathcal O}(10^2) - {\mathcal O}(10^2)$ smaller than that of KS. AKRA achieves significantly smaller L than KS for all mask cases, demonstrating enhanced localization abilities. 

To illustrate the effects of masked and unmasked pixels, we present the scatter plots of the reconstructed and true convergence ($\kappa_{\rm rec}$-$\kappa_{\rm true}$) for the unmasked and masked pixels separately in Figs.~\ref{fig:kde_obs_good} and \ref{fig:kde_obs_bad}. 
For the unmasked pixels, AKRA produces $\kappa$ maps that are almost identical to the input $\kappa$ map for all cases. In contrast, the KS method exhibits noticeable residuals even for the unmasked pixels. In the case of masked pixels, the AKRA method produces accurate $\kappa$ maps for A1 cases and demonstrates overall robustness in A2 and A3 cases, apart from specific regions with clustered masked pixels. However, the KS method exhibits substantial residuals across all cases.

\subsection{Random mask}
\label{subsec:random_mask}

The random mask is obtained by selecting pixels randomly from $m(\theta)$ and converting them to 0. The five panels in Fig. \ref{fig:random_mask} illustrate the masks with different masked pixel rates of 10\%, 20\%, 30\%, 40\%, and 50\%, respectively. 

In this section, we provide a detailed analysis of the matrix $(\mathbf{A}^{\rm{T}} \mathbf{N}^{-1} \mathbf{A})^{-1}$. This matrix has two important meanings.
First, this matrix approximates the inverse of the covariance matrix of the maximum-likelihood estimate. The matrix $\left(\mathbf{A}^{\rm{T}} \mathbf{N}^{-1} \mathbf{A}\right)^{-1}$ contains all information about the mask. However, we have made the implicit assumption that we did not account for noise. This assumption implies that we calculate the inverse covariance matrix as $\left(\mathbf{A}^{\rm{T}} \mathbf{A}\right)^{-1}$, neglecting the contribution of noise. When noise is uncorrelated across pixels, the inverse noise matrix, $\mathbf{N}^{-1}$, becomes diagonal and can also be ignored. We anticipate that the smallest covariance will be associated with the greatest inverse covariance.

Secondly, the matrix $\left(\mathbf{A}^{\rm{T}} \mathbf{N}^{-1} \mathbf{A}\right)^{-1}$ is the deconvolution matrix of the estimate in Eq.~\ref{eq:kappa_estimator}. The eigendecomposition of the deconvolution matrix is given by 
$$\left(\mathbf{A}^{\rm{T}} \mathbf{N}^{-1} \mathbf{A}\right)^{-1} = \mathbf{V} \boldsymbol{\Lambda} \mathbf{V}^{\rm{T}},$$
where $\mathbf{V}$ is a matrix whose columns are the eigenvectors and $\boldsymbol{\Lambda}$ is a diagonal matrix whose entries are the corresponding eigenvalues. The eigendecomposition can provide helpful insights into the behavior of the deconvolution matrix. For instance, the eigenvalues deermine the extent to which different components of the input signal are amplified or attenuated during the deconvolution process.

\begin{figure*}[htbp]
    \centering
    \includegraphics[width=0.7\textwidth]{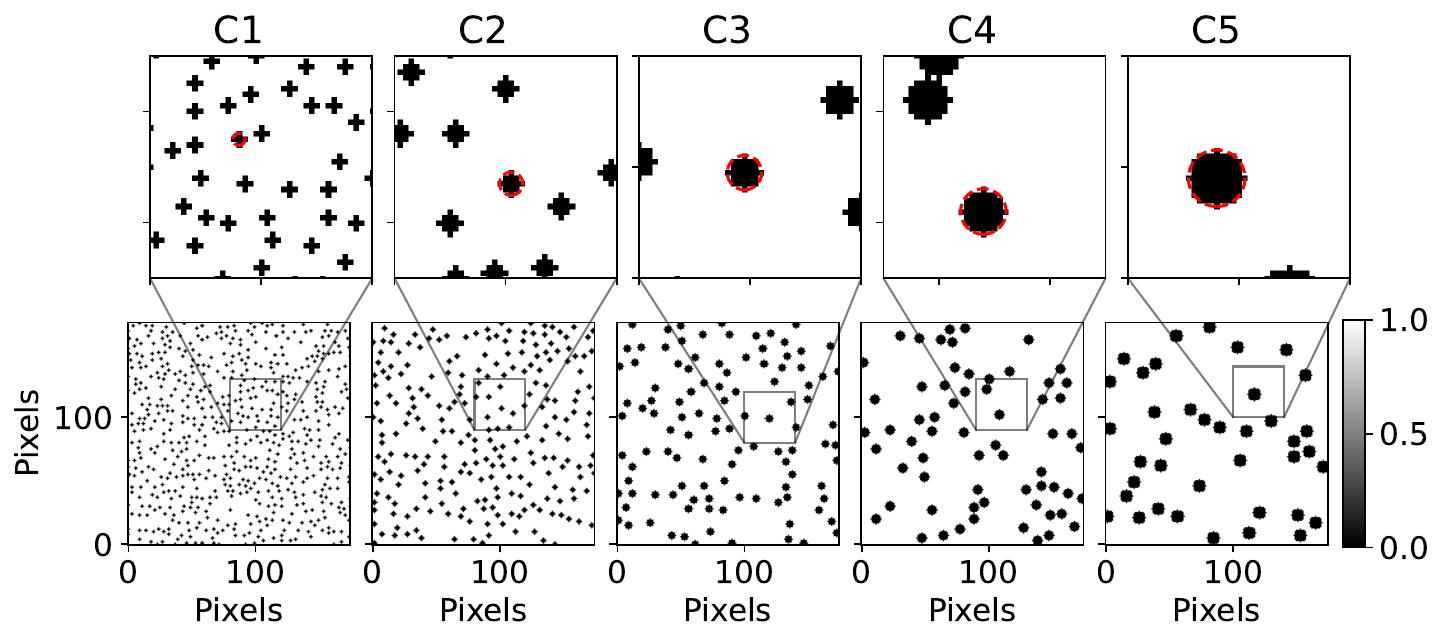}
    \caption{Circular mask with same rate (10\%) of masked pixels. From left to right, the radius of the circular mask ranges from 1 to 5 pixels, respectively. The top panels show a magnified view of the part region ($40\times 40$ pixels) in the gray rectangle. }
    \label{fig:circular_mask}
\end{figure*}

In Fig.~\ref{fig:eigenvalue}, we plotted the eigenvalue spectra of the matrix $(\mathbf{A}^{\rm{T}} \mathbf{N}^{-1} \mathbf{A})$ for five different mask maps. Our analysis of the eigenvalue spectra revealed a consistent trend: as the percentage of masked pixels increased from type B1 to B5, the amplitude of the eigenvalue spectra decreased. This decrease in amplitude indicates an increase in covariance. However, despite the variation in the eigenvalue spectra, the number of independent modes remained approximately equal across the different mask maps. This number of independent modes is determined by the two observables $\gamma_{1}^{m}(\theta)$ and $\gamma_{2}^{m}(\theta)$, which are the shear components measured within the masked region. It is worth noting that the two observables $\gamma_{1}^{m}(\theta)$ and $\gamma_{2}^{m}(\theta)$ have non-local relation with the underlying convergence field $\kappa(\theta)$ we aim to reconstruct. Therefore, in the presence of a mask with 50\% coverage, the number of independent modes is approximately equal to twice the number of unmasked pixels. This indicates that, despite the masking, there is still a substantial amount of independent information available for the reconstruction process, captured by the two observables within the unmasked region.

We also calculate the PSF matrix in Eq. \ref{eq:psf}.
Figs. \ref{fig:diag} and \ref{fig:psf_ratio} show the performance of a PSF matrix (Eq. \ref{eq:psf}) for different mask maps. A PSF matrix serves as a tool to assess the quality of $\kappa$ map reconstruction from $\gamma_1$ and $\gamma_2$ after applying a given mask. Fig. \ref{fig:diag} displays the diagonal elements of the PSF matrix, representing the self-contribution of each pixel to its reconstruction. In an ideal scenario, these values should be 1, signifying perfect reconstruction of the pixel. Fig. \ref{fig:psf_ratio} presents the ratio between the squared off-diagonal elements and the squared diagonal elements of the PSF matrix. This ratio represents the influence of neighboring pixels on the reconstruction of a given pixel. Ideally, the ratio should be 0, which indicates no cross-talk between pixels. 
As more pixels are masked, the off-diagonal elements become larger, but also considerably smaller than diagonal elements.

Fig. \ref{fig:random_mask_ps} shows a significant difference between the results obtained using AKRA and KS methods. The power spectrum of reconstructed $\kappa$ map from KS method is sensitive to the masked fraction. Conversely, the power spectrum ratio from AKRA is close to 1 for mask fraction $\leq 50\%$. This also suggests that AKRA exhibits lower sensitivity to variations in the mask fraction.
Specifically, even in the B5 case where only half of the pixels are unmasked, the power spectrum ratio and cross-correlation coefficient of AKRA remain close to 1.
The results indicate that AKRA is more effective than the KS method in reconstructing the $\kappa$ map from the shear map in the presence of a mask.

\begin{figure*}[htbp]
    \centering
    \subfigure[The residual maps normalized by r.m.s. of the true signal from AKRA method.
    ]{\includegraphics[width=0.7\textwidth]{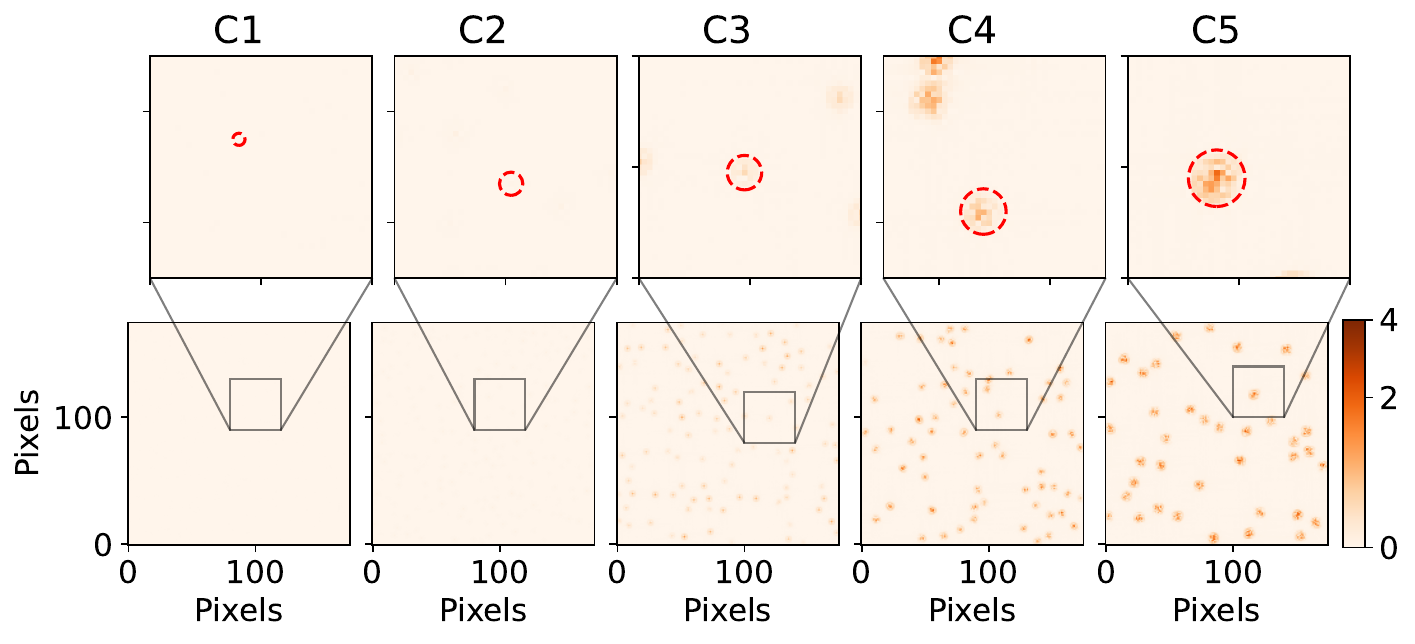}}
    \subfigure[ The residual maps normalized by r.m.s. of the true signal from KS method.  ]{\includegraphics[width=0.7\textwidth]{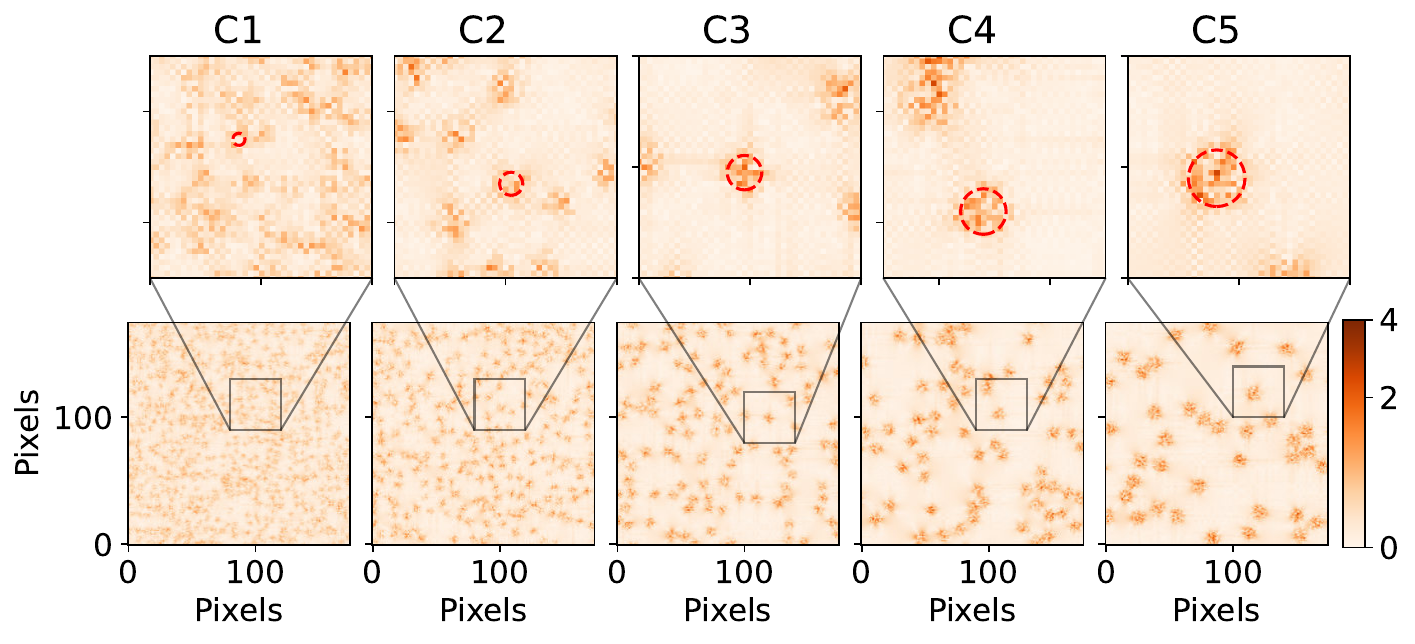}}
    \caption{The residual maps of the true $\kappa$ map and the reconstructed $\kappa$ map obtained by using AKRA and KS. The top row in each panel shows a magnified view of the central region (40 $\times$ 40 pixels) enclosed by the gray rectangle. The red circle in the magnified region indicates the boundary of the circular mask. The normalized residual from AKRA is significantly smaller than KS. In cases C1 and C2, the normalized residual approaches zero in the masked regions. However, in cases C3 to C5, where the masked pixels are more clustered, the normalized residual increases in magnitude.}
    \label{fig:circular_residual}
\end{figure*}

\begin{figure*}[htbp]
    \centering
    \includegraphics[width=0.8\textwidth]{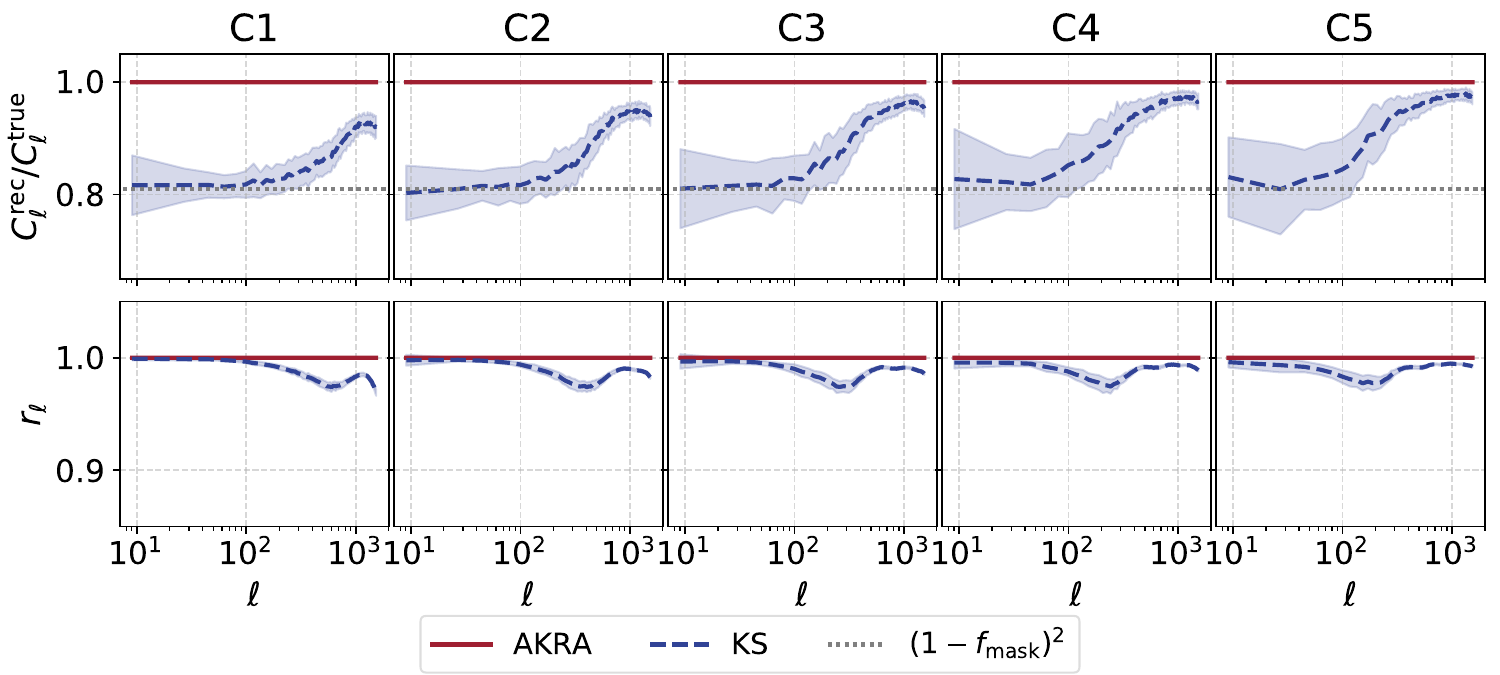}
    \caption{Same as Fig.~\ref{fig:random_mask_ps}, but for circular masks with a fixed 10\% mask fraction (unmasked pixels). The power spectrum ratio and cross-correlation coefficient from AKRA are better than that from KS. }
    \label{fig:circular_ps}
\end{figure*}

\subsection{Circular mask}
\label{subsec:circular_mask}

In this section, we use circular masks with a fixed 10\% pixel coverage to assess the performance of our algorithm. Fig.~\ref{fig:circular_mask} presents circular masks with radius ranging from 1 to 5 pixels, corresponding to cases C1 to C5, respectively.

In Fig.~\ref{fig:circular_residual}, we compare the residual maps using the KS and AKRA algorithm for circular masks with various radius. KS produces significant residual values within the masked regions. In contrast, the AKRA algorithm yields a smaller residual within these areas. Notably, for small mask radius (cases C1 and C2), the residual is reduced to zero within the masked regions. For larger radius (cases C3 to C5), the residual remains small in comparison to the residual obtained using the KS method. Accordingly, AKRA perform well at the boundary of circular mask, and successfully recovers several pixels near the masked regions.

With a constant 10\% mask fraction, AKRA invariably surpasses KS across C1-C3 cases. Strikingly, for all scenarios AKRA achieves reconstructed $C_\ell$ accurate to 1\% and better, as seen in the power spectrum ratios approaching unity. Equally impressive are the cross-correlation coefficients of $r_\ell\approx1$, laying bare near-perfect reconstructions.

Fig.~\ref{fig:circular_ps} depicts the power spectrum ratio and cross-correlation coefficient for the circular masks, when considering unmasked pixels. With a fixed masked fraction of 10\%, AKRA consistently outperforms KS. Surprisingly, it is worth mentioning that the AKRA produce $C_\ell^{\rm{rec}}$ is accurate to better than 1\% and cross correlation coefficient $r_\ell$ is close to 1 for all cases.

% With a fixed masked fraction of 10\%, AKRA consistently outperforms KS. It is worth mentioning that the results obtained using the KS method do not exhibit significant changes with varying mask radius. For C1 and C2, AKRA produces a recovered $/kappa$ map accurately, but for C3 to C5, as the mask radius increases, the power spectrum ratio and cross-correlation coefficient deviate further from 1. 

\section{Discussion and conclusion} \label{sec:result}
In this study, we present AKRA for reconstructing mass maps from weak lensing shear observations, inspired by interferometer mapmaking and CMB mapmaking techniques. Fundamentally the problem of mapmaking is one of data reduction, where for interferometers, a large set of time-ordered measurements of visibilities sample different linear combinations of the sky at various times and in different ways. Similarly, for the $\kappa$ map reconstruction, we have two observables, $\gamma_1$ and $\gamma_2$, which sample different linear combinations of the true $\kappa$ map in distinct ways. Based on it, AKRA utilizes a prior-free maximum likelihood framework to calculate the minimum variance estimate of the $\kappa$ map. Through extensive simulations with varying mask fractions (10\% to 50\%) and mask shapes, we demonstrate that AKRA not only significantly outperforms the widely-used Kaiser-Squires (KS) method in terms of map quality and summary statistics, but recovers the true convergence signal with an accuracy of $1\%$ or even better. 

Interestingly, AKRA is able to perform effectively in both unmasked and random masked pixels. In the case of unmasked pixels, AKRA achieves high accuracy, with the reconstructed auto power spectrum ($C_\ell^{\rm{rec}}$) accurate to 1\% or better, and a high cross-correlation coefficient ($r_\ell$) close to 1, even under extreme conditions of mask fraction and shape. Notably, AKRA surpasses existing reconstruction algorithms by successfully recovering the information from random masked pixels. This capability is attributed to the non-local nature of the two observables, $\gamma_1$ and $\gamma_2$, enabling effective reconstruction even in masked regions. We also introduce new ways to quantify the mask effects by examining the information content or the PSF matrix, providing insights into the impact of masks on $\kappa$ map reconstruction. The localization measure $L$ that we define further reveals to what extend the  mask in the shear catalog contaminates the convergence reconstruction in the unmasked regions due to the non-local relation between shear and convergence. $L$ of ARKA is a factor of ${\mathcal O}(10)-{\mathcal O}(10^2)$ smaller than that of KS. The typical value of $L$ in AKRA is $\lesssim 10^{-3}$, meaning that the contamination is in general negligible. This explains the excellent performance of AKRA if we only use the reconstructed convergence in the unmasked regions. 

Since in the current version we only aim to resolve the mask issue, several significant simplifications have been adopted.  Firstly, it assumes a flat sky and periodic boundary conditions, but it can be extended to handle the curved sky using the spherical harmonic transform with spin instead of the Fourier transform.  Secondly, the current implementation of AKRA assumes noise-free shear maps. This is equivalent to setting the noise covariance matrix  ${\bf N}={\bf I}$.  As a reminder, ${\bf N}\equiv \langle {\bf n}{\bf n}^T\rangle$.  In reality, the shear measurement noise is not only non-zero, but inhomogeneous, meaning a nontrivial ${\bf N}$. Since AKRA contains this key ingredient (Eq. \ref{eq:psf}), the issue of inhomogeneous measurement noise can be  appropriately considered by replacing ${\bf N}={\bf I}$ with ${\bf N}$ estimated from the realistic shear catalog.  Noise is an inherent component that can significantly affect the accuracy of mass map reconstruction. More detailed processing is required for realistic data, which includes factors such as shear measurement errors, intrinsic alignment, point spread function leakages/residuals, etc., as these elements could potentially introduce correlated noise and render the $\bf{N}$ matrix non-diagonal. Future research should focus on extending AKRA to handle noisy shear maps and properly account for the non-diagonal covariance matrix of the shear field. Extending AKRA to handle the curved sky and incorporating noise effects would be crucial for enhancing its applicability and accuracy in real observational scenarios.

\section*{Acknowledgements}
This work is supported  by National Science Foundation of China (11621303, 12273020), the National Key R\&D Program of China (2020YFC2201602), the
China Manned Space Project (\#CMS-CSST-2021-A02 \& CMS-CSST-2021-A03), and the Fundamental Research Funds for the Central Universities. This work made use of the Gravity Supercomputer at the Department of Astronomy, Shanghai Jiao Tong University. The results in this paper have been derived using the following packages: Numpy \citep{numpy}, Scipy \citep{SciPy},
\texttt{HEALPix} \citep{healpix}, IPython \citep{ipython} and CCL \citep{pyccl}.

\appendix

\begin{figure*}
    \subfigure[Same as Fig. \ref{fig:eigenvalue}, but for A1-A3 cases.]{\includegraphics[width=0.4\textwidth]{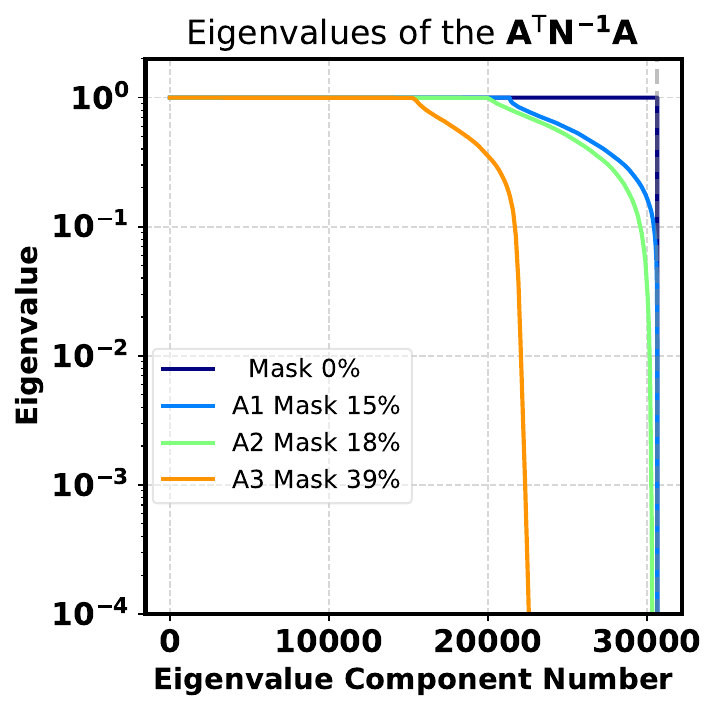}} 
    \subfigure[Same as Fig. \ref{fig:diag}, but for A1-A3 cases.]{\includegraphics[width=0.4\textwidth]{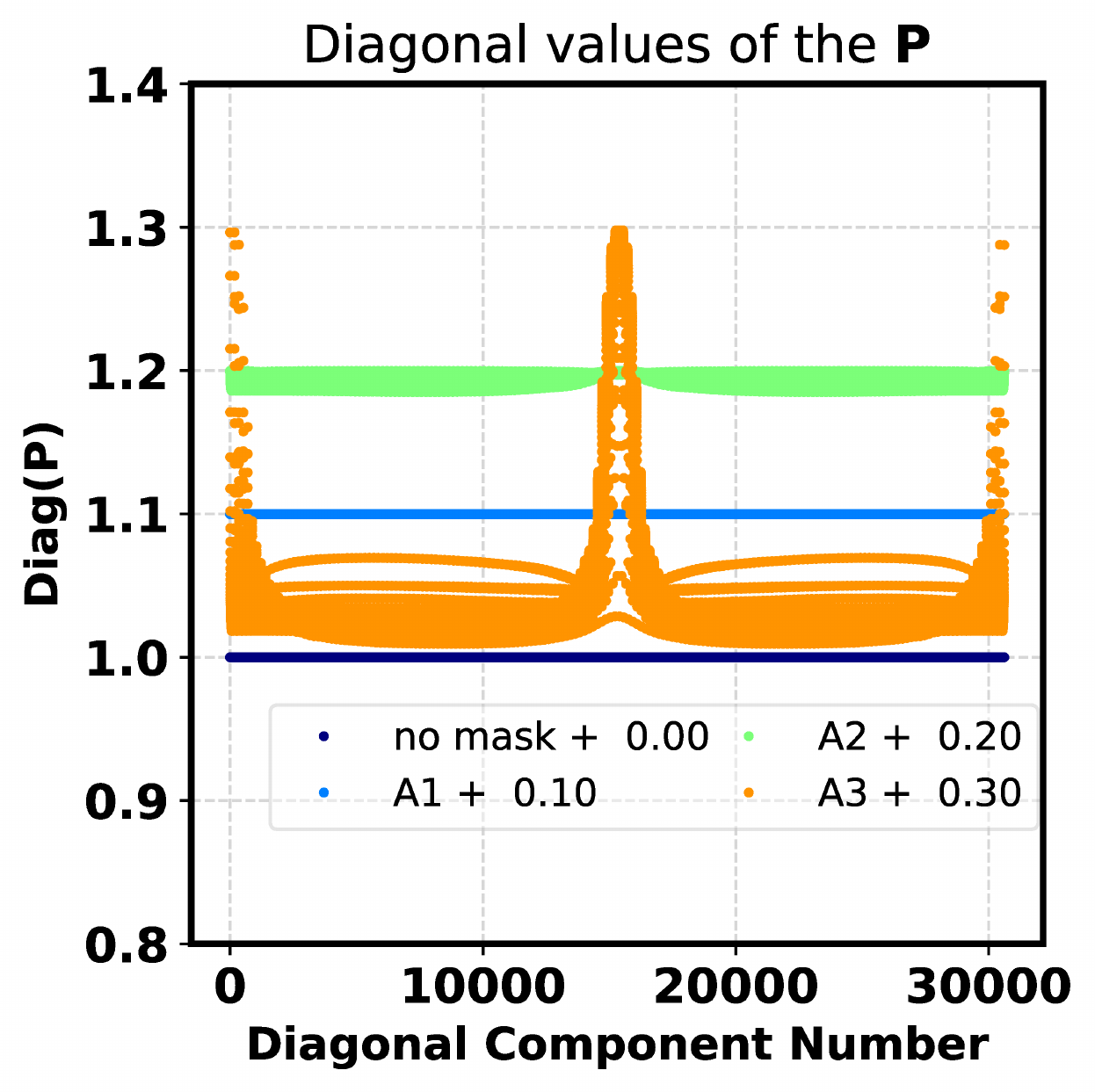}}
    \caption{Eigenvalues of $\mathbf{A}^{\rm{T}} \mathbf{N}^{-1} \mathbf{A}$ and diagonal elements of PSF matrix for A1-A3 cases.}
    \label{fig:eigenvalue_A} 
\end{figure*}

\begin{figure*}
    \subfigure[Same as Fig. \ref{fig:eigenvalue}, but for C1-C5 cases.]{\includegraphics[width=0.4\textwidth]{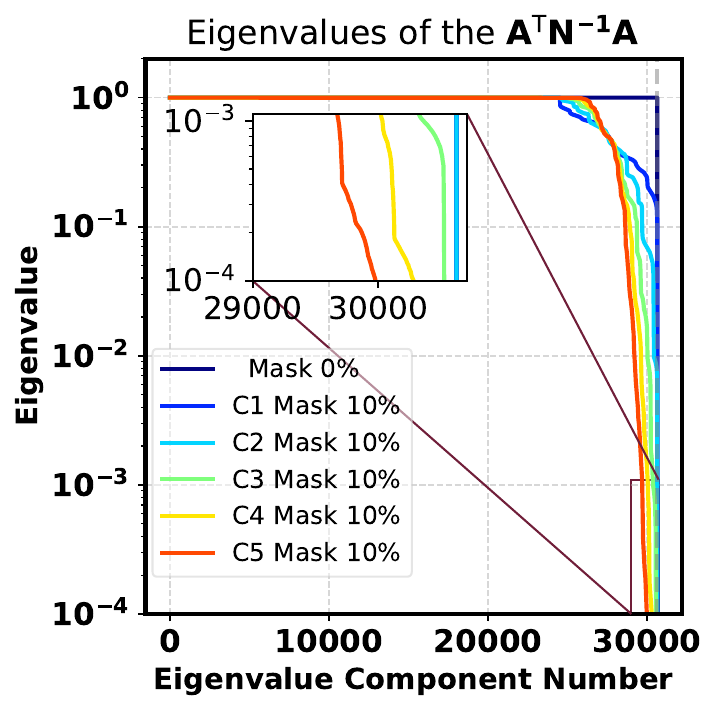}}
    \subfigure[Same as Fig. \ref{fig:diag}, but for C1-C5 cases.]{\includegraphics[width=0.4\textwidth]{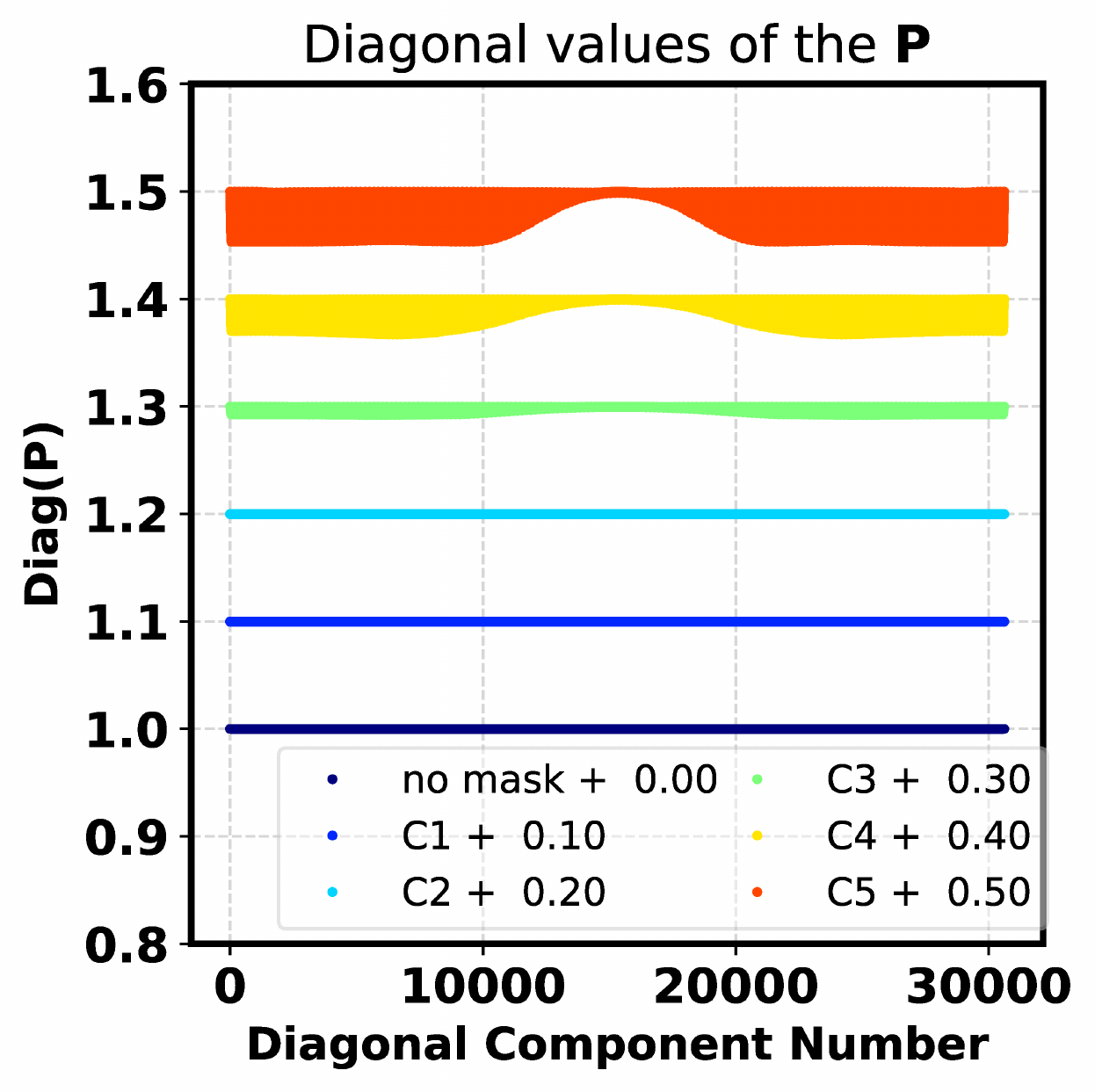}}
    \caption{Eigenvalues of $\mathbf{A}^{\rm{T}} \mathbf{N}^{-1} \mathbf{A}$ and diagonal elements of PSF matrix for C1-C5 cases.}
    \label{fig:eigenvalue_C} 
\end{figure*}

\section{Eigenvalues and PSF}
In the main text, we examined the performance of AKRA and KS for three types of masks to demonstrate AKRA's advantages. In this section, we will provide some additional results to complement the analysis in the main text.

In Sec. \ref{subsec:random_mask}, we introduced quantifying mask effects by examining the information content via the eigenvalues of the matrix $\mathbf{A}^{\rm{T}} \mathbf{N}^{-1} \mathbf{A}$ and the diagonal values of the point spread function (PSF) matrix. This provides insights into how masks impact $\kappa$ map reconstruction.

In the main text, these results were only shown for the B1-B5 mask cases. Here we present the eigenvalues of matrix $\mathbf{A}^{\rm{T}} \mathbf{N}^{-1} \mathbf{A}$ and diagonal values for A1-A3 and C1-C5 cases. 

Fig. \ref{fig:eigenvalue_A}(a) shows that the eigenvalues decrease as mask fraction increases from A1 to A2, respectively, but similar independent modes can be reconstructed. This also agrees with our finding in the main text that the inverse covariance declines with larger masks. In contrast, the irregularly shaped A3 mask yields significantly lower eigenvalues compared to A1 and A2, suggesting its geometry affects the information content more than the fractional coverage alone. Fig. \ref{fig:eigenvalue_C}(a) also indicates that clustered circular masks like C3-C5 (radius 3-5 pixels) will lose more independent modes than less clustered C1-C2 (radius 1-2 pixels) of the same mask fraction.
The eigenvalue spectra also suggest the mask geometry/clustering affects information content, not just mask fraction.

Figs. \ref{fig:eigenvalue_A}(b) and \ref{fig:eigenvalue_C}(b) also show the diagonal values of PSF matrix correlate with reconstruction quality trends. Scenarios with diagonal values closer to unity (e.g. A1, C1, C2) exhibit smaller residuals and better power spectrum/cross-correlation recovery in the main text.

\section{Results for all pixels}

\begin{figure*}
    \centering
    \includegraphics[width=0.8\textwidth]{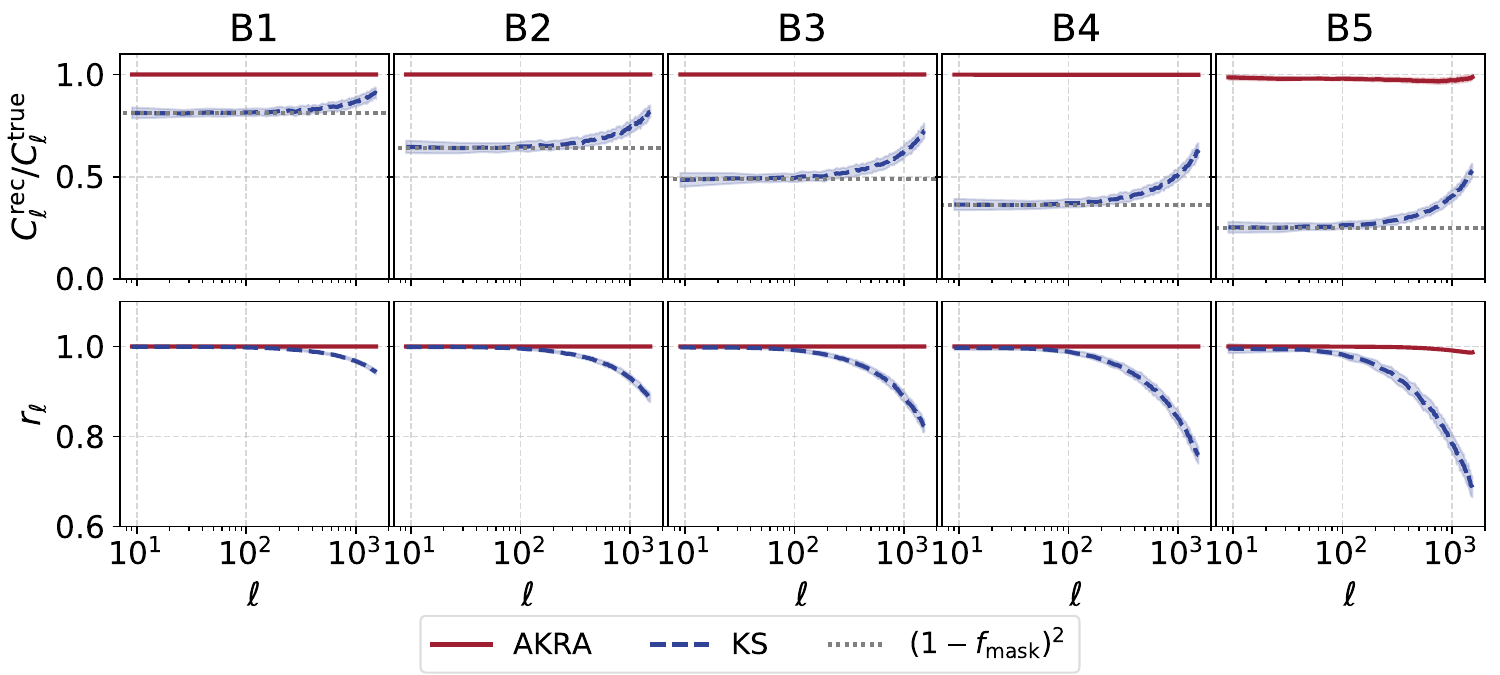}
    \caption{Same as Fig.~\ref{fig:random_mask_ps}, but for all pixels. AKRA is able to accurately recover all the pixels, with a power spectrum ratio and cross-correlation coefficient in all cases around $5\%$ and $1-r_\ell \lesssim 5\%$ respectively. }
    \label{fig:random_mask_ps_all}
\end{figure*}

\begin{figure*}
    \centering
    \includegraphics[width=0.8\textwidth]{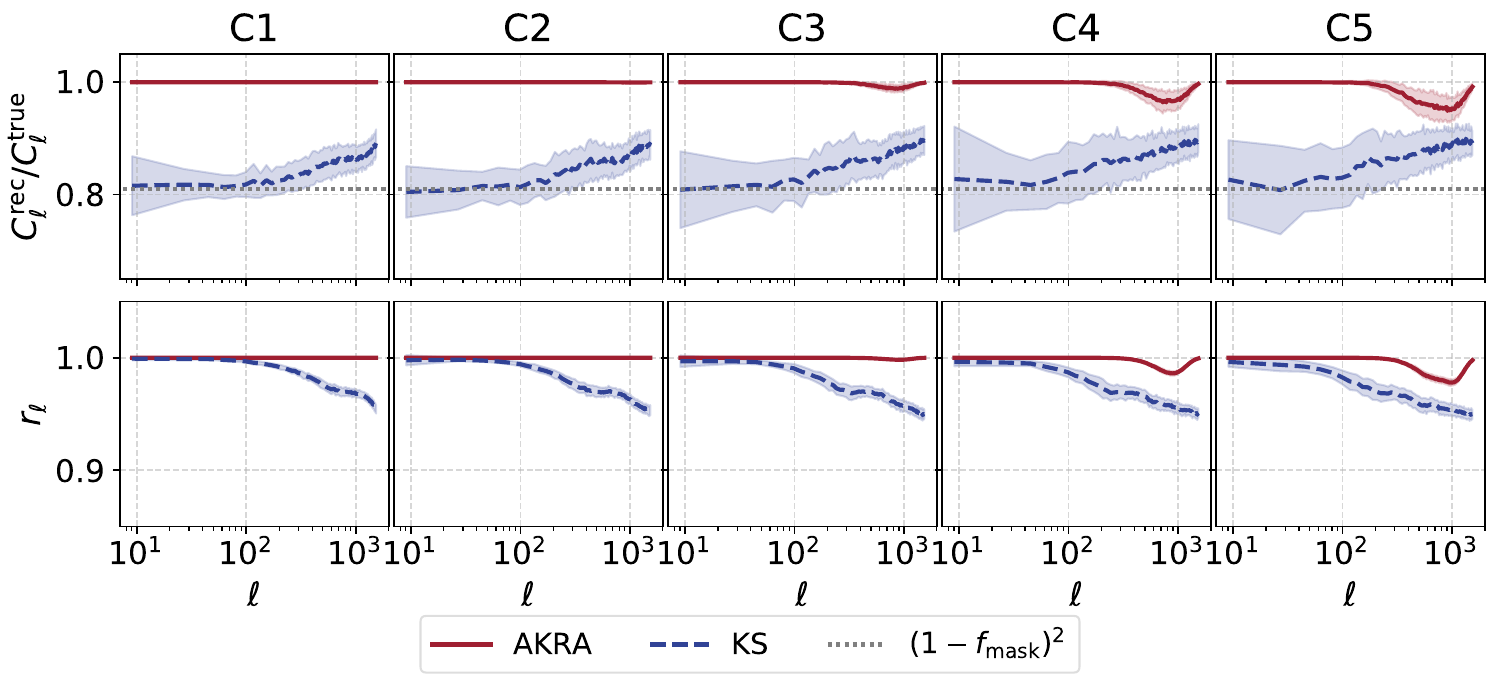}
    \caption{Same as Fig.~\ref{fig:circular_ps}, but for all pixels.  As the mask radius increases, AKRA may deviate from the ideal condition, particularly at small scales.}
    \label{fig:circular_ps_all}
\end{figure*}

We discuss the reconstruction of masked pixels in the main text. Here we complement that analysis by presenting results for AKRA when considering all pixels. Figs. \ref{fig:random_mask_ps_all} and \ref{fig:circular_ps_all} show the power spectrum ratios and cross-correlation coefficients for the B1-B5 random and C1-C5 circular masks, respectively, now over all pixels rather than just the masked regions. For the mask fraction ranging from 10\% to 50\%, AKRA achieves power spectrum ratio and coefficients within 5\% of ideal across all cases, demonstrating it can accurately recover the full original convergence field. Fig. \ref{fig:circular_ps_all} indicates AKRA maintains perfect reconstruction for C1-C2 masks. However, for larger clustered circular masks deviations up to $\lesssim 10\%$ emerge at small scales, albeit still a significant improvement over KS.

\bibliographystyle{apsrev}
\bibliography{apssamp}% Produces the bibliography via BibTeX.

\end{document}